\documentclass{aa}	
\usepackage[varg]{txfonts}

\usepackage{multirow}

\usepackage{natbib}
\bibpunct{(}{)}{;}{a}{}{,} 

\title{Effect of galaxy mergers on star formation rates}
\author{W.~J.~Pearson\inst{\ref{inst:SRON}, \ref{inst:Kapteyn}},
L.~Wang\inst{\ref{inst:SRON}, \ref{inst:Kapteyn}},
M.~Alpaslan\inst{\ref{inst:CCPP}},
I.~Baldry\inst{\ref{inst:LJHU}},
M.~Bilicki\inst{\ref{inst:PAS}},
M.~J.~I.~Brown\inst{\ref{inst:Monash}},
M.~W.~Grootes\inst{\ref{inst:eScience}},
B.~W.~Holwerda\inst{\ref{inst:Louisville}},
T.~D.~Kitching\inst{\ref{inst:Mullard}},
S.~Kruk\inst{\ref{inst:ESO}},
F.~F.~S.~van~der~Tak\inst{\ref{inst:SRON}, \ref{inst:Kapteyn}}
}
\institute{SRON Netherlands Institute for Space Research, Landleven 12, 9747 AD, Groningen, The Netherlands\label{inst:SRON}\\\email{w.j.pearson@sron.nl}
\and Kapteyn Astronomical Institute, University of Groningen, Postbus 800, 9700 AV Groningen, The Netherlands\label{inst:Kapteyn}
\and Center for Cosmology and Particle Physics, New York University, 726 Broadway, New York, NY 10012, United States\label{inst:CCPP}
\and Astrophysics Research Institute, Liverpool John Moores University, Twelve Quays House, Egerton Wharf, Birkenhead CH41 1LD, UK\label{inst:LJHU}
\and Center for Theoretical Physics, Polish Academy of Sciences, al. Lotnik\'{o}w 32/46, 02-668, Warsaw, Poland\label{inst:PAS}
\and School of Physics and Astronomy, Monash University, Clayton, Victoria 3800, Australia\label{inst:Monash}
\and Netherlands eScience Center, Science Park 140, 1098 XG Amsterdam, The Netherlands\label{inst:eScience}
\and Department of Physics and Astronomy, 102 Natural Science Building, University of Louisville, Louisville KY 40292, USA\label{inst:Louisville}
\and Mullard Space Science Laboratory, Dorking, RH5 6NT, UK\label{inst:Mullard}
\and European Space Agency, ESTEC, Keplerlaan 1, NL-2201 AZ, Noordwijk, The Netherlands.\label{inst:ESO}
}

\authorrunning{W.~J.~Pearson et al.}

\date{Received 18 July 2019 /
Accepted 27 August 2019}
\abstract{Galaxy mergers and interactions are an integral part of our basic understanding of how galaxies grow and evolve over time. However, the effect that galaxy mergers have on star formation rates (SFR) is contested, with observations of galaxy mergers showing reduced, enhanced and highly enhanced star formation.}
{We aim to determine the effect of galaxy mergers on the SFR of galaxies using statistically large samples of galaxies, totalling over 200\,000, over a large redshift range, 0.0 to 4.0.}
{We train and use convolutional neural networks to create binary merger identifications (merger or non-merger) in the SDSS, KiDS and CANDELS imaging surveys. We then compare the galaxy main sequence subtracted SFR of the merging and non-merging galaxies to determine what effect, if any, a galaxy merger has on SFR.}
{We find that the SFR of merging galaxies are not significantly different from the SFR of non-merging systems. The changes in the average SFR seen in the star forming population when a galaxy is merging are small, of the order of a factor of 1.2. However, the higher the SFR above the galaxy main sequence, the higher the fraction of galaxy mergers.}
{Galaxy mergers have little effect on the SFR of the majority of merging galaxies compared to the non-merging galaxies. The typical change in SFR is less than 0.1~dex in either direction. Larger changes in SFR can be seen but are less common. The increase in merger fraction as the distance above the galaxy main sequence increases demonstrates that galaxy mergers can induce starbursts.}


\begin{document}
\maketitle

\section{Introduction}
Galaxy mergers and interactions form a key part of our understanding of how galaxies form and evolve over time. In cold dark matter cosmology, dark matter halos merge under hierarchical growth that results in the merger of the halos' baryonic counterparts \citep[e.g.][]{2014ARA&A..52..291C, 2015ARA&A..53...51S}. This interaction results in the disruption of the galaxies that lie at the centre of the dark matter halos. Tidal forces act to pull and distort the galaxies, moving stars within the galaxies from the disk to the spheroid component \citep[e.g.][and references therein]{1972ApJ...178..623T, 2015ARA&A..53...51S}. Mergers can potentially increase the activity of an active galactic nucleus \citep[e.g.][]{1996ARA&A..34..749S, 2019MNRAS.487.2491E}, although more recent work suggests this may not always be the case \citep[e.g.][]{2010MNRAS.401.1552D, 2018MNRAS.476.2308W}.

Mergers are also thought to trigger periods of extreme star formation: starbursts. From simulations, these starbursts are believed to be a result of the tidal interactions of the galaxies compressing and shocking the gas, resulting in the rapid formation of stars \citep[e.g.][]{2004MNRAS.350..798B, 2009ApJ...694L.123K, 2009PASJ...61..481S}. Such shock-induced star formation in mergers has also been observed \citep{schweizer2009}. These intense star forming events are believed to be the cause of some of the brightest infrared objects, ultra luminous infrared galaxies \citep{1996ARA&A..34..749S, 2012MNRAS.421.1539N}. This connection between starbursts and merging galaxies resulted in the prevailing theory that most merging galaxies go through a starburst phase \citep[e.g.][]{1985MNRAS.214...87J, 2005ASSL..329..143S}.

More recent observations have shown that merger induced starbursts are found in the minority of merging systems. These studies have found that the typical increase in star formation rate (SFR) of a merger is at most a factor of two, much lower than what would typically be considered a starburst \citep{2013MNRAS.435.3627E, 2015MNRAS.454.1742K, 2018ApJ...868...46S}. Work by \citet{2015MNRAS.454.1742K} has shown that the majority of galaxy mergers are found to cause a reduction in the SFR when compared to non-merging galaxies of comparable stellar masses. In total, approximately 10-20\% of star forming galaxies are found to be undergoing a merger \citep{2014ApJ...789L..16L, 2019MNRAS.485.5631C}, with this fraction increasing with redshift \citep{2006ApJ...652...56B, 2009MNRAS.394.1956C, 2010A&A...518A..20L, 2011ApJ...742..103L, 2015A&A...576A..53L, 2017MNRAS.470..651R}. While observational evidence for the change of SFR as a function of time before and after a merger is difficult due to the long timescales involved, there is observational evidence for starbursts on the first and second close passes of two galaxies as well as coalescence. These bursts appear to last between 10$^{7}$ and 10$^{8}$ years \citep{2017A&A...607A..70C}. This is supported by observations that show close pairs have higher SFRs than more separated galaxies in mergers \citep{2015MNRAS.452..616D}.

Gas rich (wet) mergers are able to support higher SFRs as there is an abundance of fuel available to create new stars \citep[e.g.][]{2008ApJ...681..232L, 2011MNRAS.417..580P, 2016ApJ...821...90A}. Gas poor (dry) mergers, however, do not have gas readily available and so it is harder to form starbursts in these systems \citep[e.g.][]{2006ApJ...640..241B, 2006ApJ...636L..81N, 2008ApJ...681..232L}. As a result of dense environments containing a larger number of gas poor galaxies than gas rich galaxies, dry galaxy mergers dominate over wet mergers in dense environments \citep{2010ApJ...718.1158L}. The fraction of dry mergers also increases with the age of the Universe \citep{2008ApJ...681..232L}. As a result of gas poor galaxies dominating at high masses (stellar mass $\gtrsim 10^{10.7}$ M$_{\odot}$), mergers of two high mass galaxies tend to be dry and, as a result, can act to suppress star formation \citep{2014MNRAS.444.3986R}.

A study by \citet{2015MNRAS.452..616D} has found that the merger ratio of the merging galaxies also influences the SFR. In major mergers (mass ratio < 3:1), the lower mass galaxy experiences a short period of enhanced star formation while in minor mergers (mass ratio > 3:1) the star formation in the lower mass galaxy is suppressed. The more massive of the merging galaxies, however, always experiences an increase in SFR regardless of whether the merger is major or minor.

Simulations of mergers have been conducted, allowing us to study the SFRs of the merging galaxies throughout the entire merger sequence from first passage to coalescence \citep[e.g.][]{2005MNRAS.361..776S, 2006ApJS..163....1H, 2008ApJ...679.1173R, 2010ApJ...710L.156R}. These simulations have shown that SFR is enhanced when the merging galaxies are close to one another: at first pass, second pass and coalescence \citep{2019MNRAS.485.1320M}. The period between first and second passes also maintains a higher SFR than in an isolated galaxy, by approximately a factor of two. This period is the majority of the merger sequence, taking approximately 2.5~Gyr of the entire 3.5~Gyr merger timescale \citep{2019MNRAS.485.1320M}. This can explain why so few galaxies are observed to be in the starburst phase of a merger as the period between starbursts is much longer than the starburst period of approximately 0.5~Gyr. The starburst caused by the close passage and coalescence is also found to be stronger for head on collisions and reduces in strength as the approach of the galaxies becomes more oblique. However, the strength of a starburst is also connected to the resolution of the simulation, with lower resolution simulations finding weaker starbursts \citep{2016MNRAS.462.2418S}.

A major observational challenge of merger studies is the difficulty in detecting a large sample of merging galaxies. Visually identifying galaxies is time consuming and hard to reproduce; different people can classify the same galaxy differently and the same classifier may assign different labels on different days. Some of this difficulty can be reduced by employing citizen science, such as Galaxy Zoo\footnote{\url{http://www.galaxyzoo.org/}} \citep[GZ;][]{2008MNRAS.389.1179L}, to get many members of the public to classify images of galaxies. However, such approaches are not scalable to the volume of data we expect from upcoming large surveys. Using non-parametric statistics, such as concentration, asymmetry, smoothness \citep[CAS; e.g.][]{2000AJ....119.2645B, 2000ApJ...529..886C, 2001defi.conf..170W, 2003AJ....126.1183C} or the Gini coefficient, a description of the relative distribution of flux within pixels, and the second-order moment of the brightest 20\% of the light \citep[$M_{20}$;][]{2004AJ....128..163L} avoids the issues with reproducibility, especially combined with detailed galaxy merger modelling to provide a classification baseline \citep{2010MNRAS.404..590L, 2010MNRAS.404..575L}. However, merger detection with these non-parametric statistics is sensitive to image quality and resolution and suffers from a high fraction of misidentifications \citep{2015ApJS..221....8H}. The close pair method is also often employed, finding pairs of galaxies that are close on the sky and in redshift \citep[e.g.][]{2000ApJ...530..660B, 2003MNRAS.346.1189L, 2005AJ....130.1516D, 2008AJ....135.1877E, 2018MNRAS.475.5133R, 2019ApJ...876..110D}. However, this method requires highly complete spectroscopic observations and can be contaminated with flybys \citep{2012ApJ...751...17S, 2014ApJ...790L..33L}.

Deep learning has the potential to overcome some of these difficulties. Once trained, neural networks are able to perform visual like classifications of galaxies, and other astronomical objects, in a fraction of the time it takes a human, or team of humans, to classify the same objects. The classifications are also reproducible: if the same object is passed through the same neural network the result will always be the same. Deep learning techniques are becoming more commonplace in the astronomical community with uses including star-galaxy classification \citep[e.g.][]{2017MNRAS.464.4463K}, galaxy morphology classification \citep[e.g.][]{2015MNRAS.450.1441D, 2015ApJS..221....8H, 2019MNRAS.484...93D}, gravitational lens identification \citep[e.g.][]{2017MNRAS.472.1129P, 2019MNRAS.tmp.1298D} and galaxy merger identification \citep[e.g.][]{2018MNRAS.479..415A, 2019A&A...626A..49P}.

In this work we aim to use deep learning techniques to identify merging galaxies within three data sets: the Sloan Digital Sky Survey \citep[SDSS;][]{2000AJ....120.1579Y}, the Kilo Degree Survey \citep[KiDS;][]{2013Msngr.154...44D, 2013ExA....35...25D} and the Cosmic Assembly Near-infrared Deep Extragalactic Legacy Survey \citep[CANDELS;][]{2011ApJS..197...35G, 2011ApJS..197...36K}. These three data sets are employed so a large range of redshifts can be covered, with SDSS and KiDS at low redshift and CANDELS at high redshift. With these identifications, we will compare the SFRs of the star forming merging galaxies with the star forming non-merging galaxies and determine if galaxy mergers have an effect on the SFR of the merging galaxies.

The paper is structured as follows. Section \ref{sec:data} discusses the data used and the merger selection for training our neural network. Section \ref{sec:tools} describes the tools used in this study, including how we determined the galaxy main sequence through modelling and a brief description of the type of deep learning we employ: convolutional neural networks. This is followed by our results and discussion in Sects. \ref{sec:results} and \ref{sec:discussion} before we conclude in Sect. \ref{sec:conclusion}. Where necessary, Wilkinson Microwave Anisotropy Probe year 7 (WMAP7) cosmology \citep{2011ApJS..192...18K, 2011ApJS..192...16L} is adopted, with $\Omega_{M}$ = 0.272, $\Omega_{\Lambda}$ = 0.728 and H$_{0}$ = 70.4~km~s$^{-1}$~Mpc$^{-1}$.

\section{Data}\label{sec:data}
To train the neural network, a large number of images of pre-classified merging and non-merging systems are required. We also collect images of unclassified images from the same surveys to classify with our networks to increase the sample size for this study. To determine if galaxy mergers affect star formation rates in the galaxies, we also require stellar masses (M$_{\star}$), SFR and redshifts for the pre-classified and unclassified galaxies. We gather these for SDSS, KiDS and CANDELS.

These three data sets cover different redshift ranges for which merger detection is attempted: the SDSS data that we use covers the low redshift regime ($0.005 < z \leq 0.1$), along with the KiDS data ($0.00 < z \leq 0.15$), while the CANDELS data that we use goes to high redshift ($0.0 < z \leq 4.0$). The overlaps in the redshifts also allow us to examine differences due to resolution, depth and other effects, by comparing the SDSS and KiDS results, as well as different wavelengths, by comparing the optical SDSS and KiDS with the near-infrared CANDELS. The CANDELS data also probes rest frame optical data at $z \approx 1.2$ with the three CANDELS bands used (1.6~$\mu$m, 1.25~$\mu$m and 814~nm) probing approximately the rest frame i, r and g bands used in the SDSS data.

\subsection{SDSS Data Release 7}
For the SDSS data, we use the network trained on SDSS images from \citet{2019A&A...626A..49P}. The merging and non-merging galaxies used to train this network were collected following \citet{2018MNRAS.479..415A}. The 3003 merging galaxies are from \citet{2010MNRAS.401.1552D, 2010MNRAS.401.1043D}, itself derived from classifications from the Galaxy Zoo (GZ) visual classification. These galaxies have GZ merger classification greater than 0.4 and were then visually checked again to ensure these galaxies are likely to be merging pairs. Approximately half (54\%) of these merging galaxies are major mergers \citep{2010MNRAS.401.1043D}, that is the ratio of the stellar masses of the two galaxies is less than three. For the non-merging galaxies, 3003 galaxies were randomly selected from galaxies that have their GZ merger classification less than 0.2. Cut-outs of the merging and non-merging objects were then requested from the SDSS cut-out server for data release 7\footnote{\url{http://cas.sdss.org/dr7/en/tools/chart/default.asp}} (DR7) to create 6006 images in the gri bands, each of 256$\times$256 pixels and with \citet{2004PASP..116..133L} colour scaling. These images were then cropped to the central 64$\times$64 pixels, corresponding to 25.3$\times$25.3~arcsec or 46.5$\times$46.5~kpc at $z = 0.1$, to reduce memory requirements while training. The merger fraction of the complete training sample, before randomly selecting the non-merging galaxies but after mass completeness cuts, is $\sim$1.0\%.

To increase the sample for analysis, all SDSS galaxies with spectroscopic redshifts between 0.005 and 0.1 were selected, to match the redshift range of the training sample, and were then classified into merging and non-merging by the \citet{2019A&A...626A..49P} network, a total of 206\,037 galaxies once selected for mass completeness. Again, 256$\times$256 pixel cutouts in the gri bands were collected for these galaxies from the SDSS DR7 cutout server and the central 64$\times$64 pixels used for classification. The M$_{\star}$ and SFR for these objects were then collected from the MPA-JHU catalogue\footnote{\url{https://wwwmpa.mpa-garching.mpg.de/SDSS/DR7/}}, which uses the \citet{2001MNRAS.322..231K} initial mass function (IMF) \citep{2003MNRAS.341...33K, 2007ApJS..173..267S, 2004MNRAS.351.1151B}. The M$_{\star}$ is therefore derived from spectral energy distribution (SED) fitting while the SFR is derived from H$\alpha$ observations.

\begin{figure}
	\centering
	\includegraphics[width=0.5\textwidth]{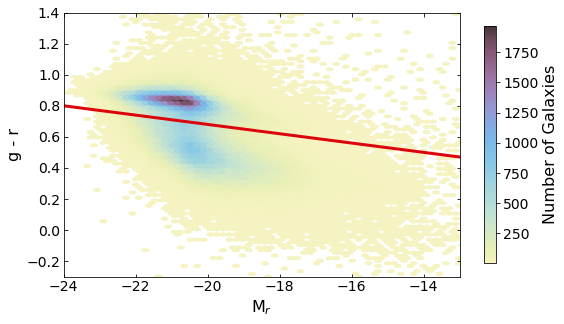}
	\caption{Rest frame $g-r$ colour vs. the absolute r magnitude (M$_{r}$) for SDSS DR7. The colour cut is shown as a red line where galaxies below the line are considered to be star forming.}
	\label{fig:sdss-colour}
\end{figure}

For determining the galaxy main sequence (MS), the star forming galaxies were selected by performing a cut in the $g-r$ - absolute r magnitude (M$_{r}$) plane, closely following \citet{2012MNRAS.420.1239L}, where we define star forming galaxies as:
\begin{equation}\label{eqn:gr}
	g - r < 0.08 - 0.03\mathrm{M}_{r}.
\end{equation}
The rest frame $g-r$ colour was determined by our own fitting of the five SDSS bands with CIGALE \citep{2009A&A...507.1793N, 2019A&A...622A.103B}. A UVJ colour cut, which is used for the KiDS and CANDELS data, is not used as the wavelength coverage is not sufficient to reliably constrain the J band magnitude. A plot of this colour cut can be seen in Fig. \ref{fig:sdss-colour}. The mass limit was determined to be log(M$_{\star}$/M$_{\odot}$) = 10.1, see Sect. \ref{subsec:mass} for details.

\subsection{KiDS}
For our KiDS sample, we use the latest data release 4 \citep[DR4;][]{2019A&A...625A...2K}. We match these catalogues with the Galaxy and Mass assembly \citep[GAMA;][]{2009A&G....50e..12D} GZ catalogue \citep[][Kelvin et al. in prep.]{Holwerda} to determine the merging and non-merging galaxies and combine this classification with non-parametric statistics (see Sect. \ref{sec:kids-mergers}). For the KiDS data, we only use r-band images to train the network, using 64$\times$64 pixel cutouts, corresponding to 13.7$\times$13.7~arcsec or 25.2$\times$25.2~kpc at $z = 0.1$, and with linear colour scaling. Tests comparing multi-channel, as used with SDSS and CANDELS, and single channel images, as used with KiDS, to identify galaxy mergers have shown that using a single channel does not notably affect the results. When applying the trained CNN to unclassified objects, we use objects that lie within the GAMA09 field. This region is large enough to provide a statistically significant sample size of galaxies and has the added benefit that it has \textit{Herschel} Spectral and Photometric Imaging Receiver \citep[SPIRE;][]{2010AA...518L...3G} coverage to aid with determining SFRs.

\begin{figure}
	\centering
	\includegraphics[width=0.5\textwidth]{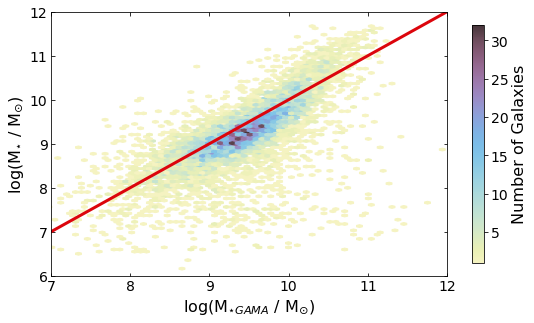}
	\caption{Comparison of the M$_{\star}$ from this work (y-axis) with M$_{\star}$ from the GAMA survey (x-axis). The red line denotes the 1-to-1 relation. The two data sets are in reasonable agreement with the average stellar masses within 0.2~dex and remain the same with and without the inclusion of SPIRE data. The typical statistical error on M$_{\star}$ is 0.1~dex.}
	\label{fig:KiDSGAMA-mass}
\end{figure}

\begin{figure}
	\centering
	\includegraphics[width=0.5\textwidth]{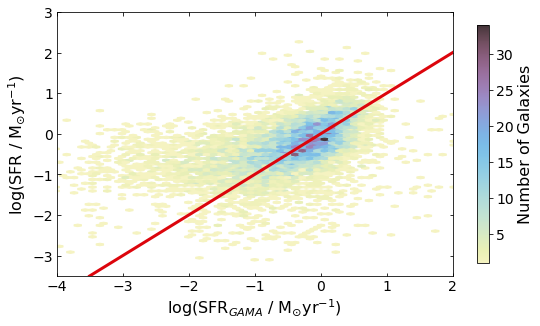}
	\caption{Comparison of the SFR from this work (y-axis) with SFR from the GAMA survey (x-axis). The red line denotes the 1-to-1 relation. The two data sets are within 0.2~dex on average and are consistent within the typical error of 0.26~dex. Both the GAMA SFRs and the SFRs from this work are derived from SED fitting.}
	\label{fig:KiDSGAMA-sfr}
\end{figure}

The majority of the KiDS objects in DR4 do not have estimates of physical parameters, beyond photometric redshifts \citep{2019A&A...625A...2K}. Thus, to derive M$_{\star}$ and SFR, we use the 9-band catalogues combined with \textit{Herschel} ATLAS \citep{2010PASP..122..499E, 2017ApJS..233...26S} SPIRE data de-blended with XID+ \citep[][see also Appendix \ref{app:cigale}]{2017MNRAS.464..885H, 2017A&A...603A.102P}. From the 9-band catalogue we use the KiDS Gaussian	aperture and point spread function \citep[GAAP;][]{2015MNRAS.454.3500K} flux densities for the ugri optical bands and the VISTA Kilo-Degree Infrared Galaxy Survey \citep[VIKING;][]{2013Msngr.154...32E} GAAP flux densities for the ZYJHKs bands, all left uncorrected for foreground extinction. SEDs are fitted to these data using CIGALE and stellar populations with a \citet{2003PASP..115..763C} IMF. As can be seen in Fig. \ref{fig:KiDSGAMA-mass}, the M$_{\star}$ from CIGALE are in good agreement, within 0.2~dex on average, with those from the GAMA survey \citep{2017MNRAS.470..283W} estimated using the MAGPHYS \citep{2008MNRAS.388.1595D} SED fitting tool, which also uses the \citet{2003PASP..115..763C} IMF. A similar comparison is made with the SFR in Fig \ref{fig:KiDSGAMA-sfr}, showing good agreement with GAMA.

\begin{figure}
	\centering
	\includegraphics[width=0.5\textwidth]{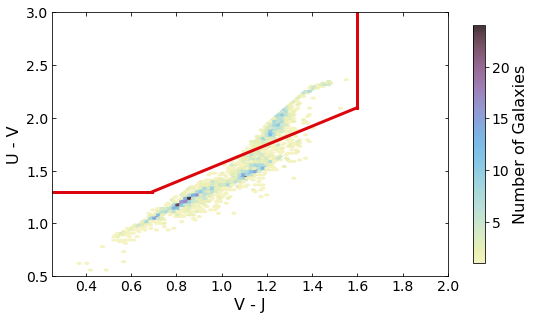}
	\caption{Rest frame U-V colour vs. rest frame V-J colour for KiDS-z00 $(0.00 < z \leq 0.15$). The colour cut is shown as a red line where galaxies below and to the right of the line are considered to be star forming.}
	\label{fig:kids-colour}
\end{figure}

To select the star forming galaxies for determining the MS, a UVJ colour cut was employed using the rest frame U-V and V-J colours, determined by CIGALE during the fitting to estimate M$_{\star}$ and SFR, and the photometric redshifts. For this, we follow \citet{2011ApJ...735...86W}:
\begin{equation}\label{eqn:UVJ}
	\begin{aligned}
		(U - V) > 0.88 \times (V - J) + 0.69 && z < 0.5\\
		(U - V) > 0.88 \times (V - J) + 0.59 && z > 0.5\\
		(U - V) > 1.3, (V - J) < 1.6 && z < 1.5\\
		(U - V) > 1.3, (V - J) < 1.5 && 1.5 < z < 2.0\\
		(U - V) > 1.2, (V - J) < 1.4 && 2.0 < z < 4.0\\
	\end{aligned}
\end{equation}
where any galaxies that do not meet these criteria are determined to be star forming. An example of the colour cut is shown in Fig. \ref{fig:kids-colour}. The mass completeness limit for the KiDS galaxies was determined to be log(M$_{\star}$/M$_{\odot}$) = 9.6, see Sect. \ref{subsec:mass} for details. This was determined using the magnitude limit from the GAMA survey of 19.8, for the r-band, as this is the limit imposed on the training sample.

\subsubsection{KiDS-GAMA Galaxy Zoo}
There are no pre-existing merger catalogues for the KiDS survey, although there are visual GZ classifications for 36\,706 galaxies in the regions that overlap with the GAMA survey (KiDS-GAMA): the three GAMA equatorial fields. We can use this classification to help select a sample of merging galaxies to use with the KiDS data. As with other Galaxy Zoo \citep{2008MNRAS.389.1179L} projects, citizen scientists were asked to classify images of galaxies following a classification tree, as described in \citet{Holwerda}, through the GZ web interface\footnote{\url{http://www.galaxyzoo.org/}} and we use the vote fractions that are weighted for user performance. These weighted vote fractions have votes from users that frequently disagree with the majority of other users weighted lower, reducing their influence on the overall vote fraction. These galaxies were selected to have redshifts between 0.002 and 0.15 and GAMA data quality flags are used to ensure only science targets are shown. Of interest here is the question concerning galaxy interactions. This question asks the classifier to identify merging galaxies, galaxies with tidal tails, galaxies that are both merging and have tidal tails or galaxies that show neither of these features. The latter of these classifications, galaxies that have neither tidal tails nor show evidence of a merger, is what will be used to help identify galaxy mergers and will hence forth be referred to as \texttt{merger\_neither\_frac}. Galaxies that have \texttt{merger\_neither\_frac} less than 0.5, that is less than half the people who classified the galaxy thought it showed no tidal features or merger indications, will be used here to for the basis of the merging galaxy sample with further refinements added.

\subsubsection{KiDS merger selection}\label{sec:kids-mergers}
The visual GZ merger classifications require validation with other methods, as chance projections or star-galaxy overlaps can be misidentified as merging galaxies \citep{2010MNRAS.401.1552D, 2010MNRAS.401.1043D}. To do this, we use the Gini, the second-order moment of the brightest 20 percent of the light ($M_{20}$), concentration (C), asymmetry (A) and smoothness (S) non-parametric parameters \citep{2004AJ....128..163L, 2000AJ....119.2645B, 2000ApJ...529..886C, 2001defi.conf..170W, 2003AJ....126.1183C}. For each of the galaxies in the KiDS-GAMA sample, we derive these five non-parametric statistics using the python code \texttt{statmorph} \citep{2019MNRAS.483.4140R}.

\begin{figure}
	\centering
	\includegraphics[width=0.5\textwidth]{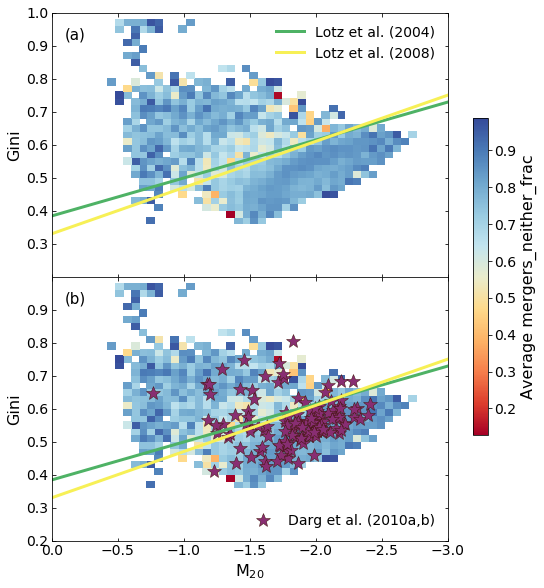}
	\caption{Gini vs. $M_{20}$ for the KiDS-GAMA GZ galaxies binned by Gini and $M_{20}$. The average \texttt{merger\_neither\_frac} from GZ within each bin is shown from low (red) to high (blue). The green line is the \citet{2004AJ....128..163L} split between merging and non-merging galaxies while the yellow line is the \citet{2008ApJ...672..177L} split. Regions with low \texttt{merger\_neither\_frac} are visually identified as merging galaxies. Panel (b) includes the visually confirmed mergers from \citet{2010MNRAS.401.1552D, 2010MNRAS.401.1043D} as purple stars.}
	\label{fig:GiniM20-notmerger}
\end{figure}

There has been found to be a division between merging and non-merging galaxies using the Gini and $M_{20}$ statistics: \citet{2004AJ....128..163L} found that galaxies can be considered to be non-mergers if
\begin{equation}\label{eqn:GiniM20L04}
	\mathrm{Gini} < -0.115 M_{20} + 0.384
\end{equation}
while \citet{2008ApJ...672..177L} found a similar result with non-mergers defined as 
\begin{equation}\label{eqn:GiniM20L08}
	\mathrm{Gini} < -0.15 M_{20} + 0.33.
\end{equation}
We also populate the Gini-$M_{20}$ parameter space, bin by Gini and $M_{20}$ and show the average \texttt{merger\_neither\_frac} of the galaxies inside each bin, as seen in Fig. \ref{fig:GiniM20-notmerger}. The \texttt{merger\_neither\_frac} is the fraction of GZ votes that say the galaxy has no indication of a galaxy merger or tidal tails. In doing this, we find that galaxies found to be mergers in the KiDS-GAMA GZ typically lie on or above these two lines. However, as can be seen in Fig. \ref{fig:GiniM20-notmerger}, there are also a large number of galaxies that lie above these lines that are classified by GZ as non-mergers: the merging galaxies appear to form a valley. Overlaying the visually confirmed merging galaxies from \citet{2010MNRAS.401.1552D, 2010MNRAS.401.1043D} that fall within the KiDS coverage, we also find that the majority of these galaxies lie below the \citet{2004AJ....128..163L} and \citet{2008ApJ...672..177L} lines, as can be seen in Fig. \ref{fig:GiniM20-notmerger}b, suggesting that this is a poor choice to determine merger status for this KiDS data set.

This disparity may be a result of the different data used. The Gini and $M_{20}$ statistics are calculated from the images and so depend on the resolution and signal-to-noise of the images \citep{2004AJ....128..163L}. The flux distribution of a lower resolution image will be different, the same flux will be spread across fewer pixels in a lower resolution images as well as removing smaller scale structures, which will increase the uncertainties in these statistics. Similarly, higher signal-to-noise images will reveal fainter features of a galaxy that will also affect the Gini and $M_{20}$. The Gini and $M_{20}$ have been found to be reasonably consistent when the signal-to-noise is above 2 but $M_{20}$ is particularly sensitive to resolution \citep{2004AJ....128..163L}. The data used in \citet{2004AJ....128..163L} is lower resolution than KiDS while \citet{2008ApJ...672..177L} uses Hubble Space Telescope data with higher resolution.

\begin{figure}
	\centering
	\includegraphics[width=0.5\textwidth]{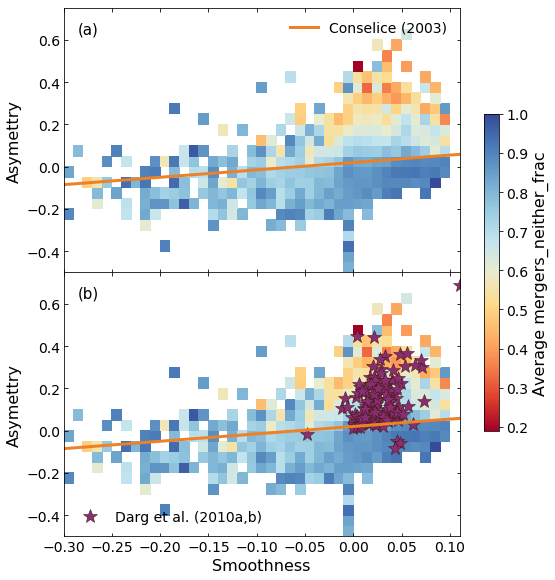}
	\caption{Asymmetry (A) vs. Smoothness (S) for the KiDS-GAMA GZ galaxies binned by A and S. The average \texttt{merger\_neither\_frac} from GZ within each bin is shown from low (red) to high (blue). Regions with low \texttt{merger\_neither\_frac} are visually identified as merging galaxies. The orange line denotes the \citet{2003ApJS..147....1C} split between merging and non-merging galaxies. Panel (b) includes the visually confirmed mergers from \citet{2010MNRAS.401.1552D, 2010MNRAS.401.1043D} as purple stars.}
	\label{fig:AS-notmerger}
\end{figure}

If instead we use the asymmetry (A) and smoothness (S) statistics, which have been found to be not overly sensitive to resolution as $M_{20}$ \citep{2004AJ....128..163L}, we find a merging sample that agrees much better with the visual classification. It has been found, by \citet{2003ApJS..147....1C}, that the merging galaxies lie above
\begin{equation}\label{eqn:AS}
	A = 0.35S + 0.02.
\end{equation}
As can be seen in Fig. \ref{fig:AS-notmerger}, this classification is in good agreement with the visual classifications from GZ. Overlaying the \citet{2010MNRAS.401.1552D, 2010MNRAS.401.1043D} mergers, we find that the majority lie above Eq. \ref{eqn:AS}. Based on this agreement, we select our merging sample to be those galaxies that have \texttt{merger\_neither\_frac} from GZ less than 0.5 and lie above Eq. \ref{eqn:AS}, with non-merging galaxies defined as those with \texttt{merger\_neither\_frac} greater than 0.5 and lie below Eq. \ref{eqn:AS}. This results in 1917 merging galaxies that we use to train the KiDS network. By matching these galaxies to the nearest galaxy within 3~arcsec in the full GAMA catalogue \citep{2017MNRAS.470..283W} and selecting the pairs that have redshifts within 0.05, we find that approximately half of these galaxies (6 of 14) are major mergers. The total number of matched pairs is very low, and will miss pairs where the secondary galaxy is below the magnitude limit of the survey, but this fraction is in line with that seen by \citet{2010MNRAS.401.1043D} in the SDSS data. We randomly select a further 1917 galaxies from the 20\,842 that lie below Eq. \ref{eqn:AS} and have \texttt{merger\_neither\_frac} greater than 0.5 to form the non-merging sample. With these classifications for merging and non-merging galaxies, and after mass completeness cuts, the merger fraction of the GZ galaxies is 8.4\%.

\subsection{CANDELS}
To train the CANDELS network, we use the visual classifications for the Great Observatories Origins Deep Survey - South \citep[GOODS-S;][]{2004ApJ...600L..93G} from \citet{2015ApJS..221...11K}. This catalogue contains galaxies with H magnitude less than 24.5 that have been classified by a small number of professional astronomers and we select objects with photometric redshift below 4.0. Of interest to this work are the classifications that identify mergers (\texttt{merger}), interaction within a segmentation map (\texttt{Int1}), interaction with a galaxy outside of the segmentation map (\texttt{Int2}), a non interacting companion (\texttt{Comp}) or no interaction (\texttt{NoInt}). During the classification, only one of these identifications may be chosen. The catalogue also contains an \texttt{Any\_Int} category, which combines the \texttt{merger}, \texttt{Int1} and \texttt{Int2} identifications. 

We define galaxies as merging if the \texttt{Any\_Int} classification is greater than 0.6 (i.e. more that 60\% of people believe that the galaxy is interacting) and we define galaxies as non-merging if the \texttt{Any\_Int} classification is less than 0.5. As with the KiDS galaxies, we match the merging galaxies to the rest of the CANDELS catalogue within 3~arcsec and selecting the pairs that have redshifts within 0.05, we find that approximately half of these galaxies (4 of 9) are major mergers. Again, the total number of matched pairs is very low, and this method will miss pairs where the secondary galaxy is below the magnitude limit of the survey, but this fraction is in line with that seen in the SDSS data. Cutouts for these objects were created from the 1.6~$\mu$m, 1.25~$\mu$m and 814~nm images. As the 814~nm images are twice the angular resolution of the other two bands, these images are reduced in size by averaging the flux density in 2$\times$2 pixel groups. The 1.6~$\mu$m, 1.25~$\mu$m and 814~nm bands are then used as the red, green and blue channels in the images, with simple linear colour scaling. As with the SDSS and KiDS images, the CANDELS images are 64$\times$64~pixels, corresponding to 3.8$\times$3.8~arcsec or 32.7$\times$32.7~kpc at $z = 1.5$. Objects with clear artefacts within the image were removed. This resulted in 694 merging galaxies and we randomly select a further 694 from the 4428 non-merging galaxies that meet our criteria. The merger fraction for the training sample using these criteria, and after mass completeness cuts, is 15.5\%.

To increase the CANDELS sample for our analysis, we classified all CANDELS galaxies with H-magnitude < 24.5 and redshift between 0.0 and 4.0, to match the training sample, from the Cosmic Evolution Survey \citep[COSMOS;][]{2007ApJS..172....1S}, Extended Groth Strip \citep[EGS;][]{2007ApJ...660L...1D}, UKIRT Infrared Deep Sky Survey (UKIDSS) Ultra-Deep Survey \citep[UDS;][]{2007MNRAS.379.1599L, 2007MNRAS.380..585C} fields with the CANDELS network (once trained). Images for these galaxies were created as above. The H-magnitude and SED derived SFR and M$_{\star}$, assuming a \citet{2003PASP..115..763C} IMF, come from \citet{2013ApJS..207...24G, 2015ApJ...801...97S} for GOODS-S, \citet{2017ApJS..228....7N} for COSMOS, \citet{2017ApJS..229...32S} for EGS and \citet{2015ApJ...801...97S} for UDS. As these catalogues contain a number of different M$_{\star}$ and SFR values, the `M\_med' is used for M$_{\star}$ and we average `SFR\_11a\_tau',   `SFR\_13a\_tau',   `SFR\_2a\_tau', `SFR\_14a', `SFR\_14a\_const', `SFR\_14a\_deltau', `SFR\_14a\_lin', `SFR\_14a\_tau', `SFR\_6a\_deltau', `SFR\_6a\_invtau' and `SFR\_6a\_tau' for SFR, as these columns are common across all catalogues. These different SFR values assume different star formation histories (SFH) where `cons' is a constant SFH, `tau' is an exponentially declining SFH, `deltau' is a delayed-exponential, `lin' is linearly increasing and `inctau' is exponentially increasing. The numbers refer to the investigator within the CANDELS team who lead the determination of that SFR \citep{2017ApJS..229...32S}. For the redshift, the `z\_best' value in the catalogues were used. This value is the spectroscopic redshift, if available, or the best photometric redshift from six members within the CANDELS team \citep{2013ApJS..207...24G, 2015ApJ...801...97S, 2017ApJS..228....7N, 2017ApJS..229...32S}

\begin{figure}
	\centering
	\includegraphics[width=0.5\textwidth]{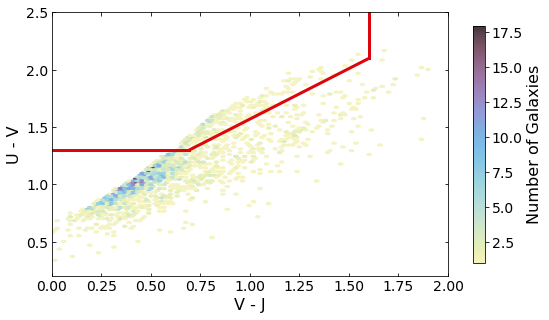}
	\caption{Rest frame $U-V$ colour vs. rest frame $V-J$ colour for CANDELS-z000. The colour cut is shown as a red line where galaxies below and to the right of the line are considered to be star forming.}
	\label{fig:candels-colour}
\end{figure}

To determine which CANDELS galaxies are star forming, we again apply the UVJ colour cuts defined in Eq. \ref{eqn:UVJ} using the rest frame U-V and V-J colours in the CANDELS catalogues, as shown in Fig. \ref{fig:candels-colour}. Mass completeness limits were calculated to be log(M$_{\star}$/M$_{\odot}$) = 8.3, 8.7, 9.1, 9.4 and 9.9 within redshift bins with edges at $z =$ 0.0, 0.6, 0.85, 1.21, 1.66 and 4.0, see also Sect. \ref{subsec:mass} below. These redshift bins were selected so there are approximately 2000 galaxies within each bin after cutting for mass completeness. For ease of reference, these redshift bins shall be referred to as CANDELS-z000, CANDELS-z060, CANDELS-z085, CANDELS-z121 and CANDELS-z166. A summary of all data sets is presented in Table \ref{table:data-summary}.

\begin{table*}
	\caption{Summary of the data used. The SDSS and KiDS limiting magnitudes are in r-band while the CANDELS limiting magnitude in H-band.}
	\begin{center}
		\begin{tabular}{ccccccc}
		\hline
		Data & Resolution & Magnitude limit & Redshift range & Mass limit & Training sample & Complete sample \\
		 & (arcsec) & & & log(M$_{\star}$/M$_{\odot}$) & per class  & (Mass limited)\\
		\hline
		SDSS & 1.4 & 17.77 & $0.005 < z \leq 0.1$ & 10.1 & 3003 & 206\,037\\
		\hline
		KiDS & 0.77 &       19.8\tablefootmark{a} & $0.00 < z \leq 0.15$ & 9.6 & 1917 & 1270 \\
		\hline
		CANDELS-z000 & 0.15 & 24.5 & $0.00 < z \leq 0.60$ & 8.3 & 694\tablefootmark{b} & 2072 \\
		CANDELS-z060 & 0.15 & 24.5 & $0.60 < z \leq 0.85$ & 8.7 & 694\tablefootmark{b} & 2004 \\
		CANDELS-z085 & 0.15 & 24.5 & $0.85 < z \leq 1.21$ & 9.1 & 694\tablefootmark{b} & 2031 \\
		CANDELS-z121 & 0.15 & 24.5 & $1.21 < z \leq 1.66$ & 9.4 & 694\tablefootmark{b} & 2010\\
		CANDELS-z166 & 0.15 & 24.5 & $1.66 < z \leq 4.00$ & 9.9 & 694\tablefootmark{b} & 1910 \\
		\hline
		\end{tabular}
	\label{table:data-summary}
	\tablefoot{
		\tablefoottext{a}{As the training set is derived from GAMA classifications, the limiting magnitude is that of the GAMA survey not that of the KiDS survey.}
		\tablefoottext{b}{The CANDELS network was trained with 694 galaxies per class for galaxies with $0.00 < z \leq 4.00$. The galaxies were split into the redshift bins shown after classification.}
	}
	\end{center}
\end{table*}

\subsection{Mass completeness}\label{subsec:mass}
Mass completeness limits were determined empirically by following \citet{2010A&A...523A..13P} and using the galaxies identified as star forming. For each galaxy, the mass the galaxy would need to have to be detected at the magnitude limit (M$_{\mathrm{lim}}$) was calculated with
\begin{equation}
	\log(\mathrm{M}_{\mathrm{lim}}) = \log(\mathrm{M}) - 0.4(x_{\mathrm{lim}} - x),
\end{equation}
where $x$ is the observed magnitude in the r-band (for SDSS and KiDS) or H-band (for CANDELS) and $x_{\mathrm{lim}}$ is the limiting magnitude of the observation. The limiting magnitudes for SDSS and CANDELS are 17.77 and 24.50 respectively. The KiDS limiting magnitude is 19.8, the limit of the GAMA survey. The faintest 20\% of objects were selected and the limiting mass was the M$_{\mathrm{lim}}$ value that 90\% of these faintest objects lie below. This was done as a function of redshift by binning the galaxies into redshift bins as described in Sect. \ref{sec:results} below. These completeness limits were then applied to the entire galaxy population.

\section{Tools}\label{sec:tools}

\subsection{Convolutional neural networks}\label{subsec:cnn}
Convolutional neural networks (CNNs) are a subset of deep learning \citep[e.g.][and references therein]{2015Natur.521..436L}. CNNs are used for image classifications and employ a series of non-linear mathematical functions, known as neurons, each with a weight and bias value. The structure of a CNN is built from a number of layers of these neurons. The lower layers are created from two-dimensional kernels that are convolved with the output of the layer below, giving CNN its name. Upper layers are one-dimensional and each neuron in these layers is connected to every neuron in the layer below. Forming a network in such a way can rapidly create a large number of neurons that require training resulting in many more free parameters within the network than there are data to train them. To reduce this dimensionality, pooling layers are employed between the lower convolutional layers. These pooling layers group the inputs into it and pass on the maximum or average value of the group, depending on the type of pooling used, with the grouping done in two-dimensions. The result is an output that is smaller in the width-height plane but has the same depth as the input. The weights and biases of the neurons within a network are trained, in the case of supervised learning used here, by passing labelled data through the network and requiring the output classification to converge on these labels. A complete and thorough description of CNNs is beyond the scope of this paper but further details are explained in \citet{lecun1998cnn}.

This paper uses the definitions of \citet{2019A&A...626A..49P} for the terms to describe the properties of CNNs. These terms may be an alternate nomenclature to other works or may be unfamiliar. To avoid confusion we reproduce these definitions in Appendix \ref{app:definitions}.

\subsubsection{Architecture of the CNN}\label{subsec:arch}
\begin{figure*}
	\centering
	\includegraphics[width=1.0\textwidth]{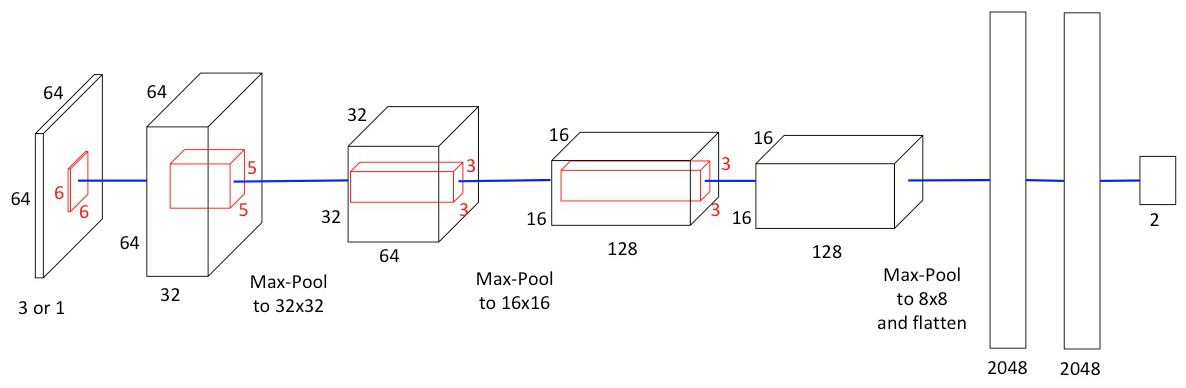}
	\caption{A schematic representation of the architecture of the CNN used with the input (three or one colour, 64$\times$64~pixel image) on the left and output (binary classification of merger or non-merger) on the right. The sizes of the kernels (red) and layers are shown with the input layer having a depth of 3 for the SDSS (gri) and CANDELS (1.6~$\mu$m, 1.25~$\mu$m and 814~nm) data and 1 for the KiDS (r-band) data.}
	\label{fig:cnn}
\end{figure*}

For this work, we use the architecture developed in \citet{2019A&A...626A..49P} and use this to train on data from CANDELS and KiDS (the network trained in \citet{2019A&A...626A..49P} is used on the SDSS images). This network is built with \texttt{Tensorflow} \citep{tensorflow2015-whitepaper} and comprises of a series of four, two-dimensional convolutional layers followed by two one-dimensional, fully connected layers of 2048 neurons. The convolutional layers have 32, 64, 128 and 128 kernels of 6$\times$6, 5$\times$5, 3$\times$3 and 3$\times$3 pixels for the first, second, third and fourth layers respectively with the stride, how far the kernel is moved as it scans the input, set at 1 pixel for all layers and the zero padding is set to ``same'' to pad each edge of the image with zeros evenly (if required). 2$\times$2 pixel max-pooling is applied after the first, second and fourth convolutional layers to reduce the dimensionality of the network. Batch normalisation \citep{ioffe2015batch} is applied after each layer, scaling the output between zero and one, and we use Rectified Linear Units \citep[ReLU;][]{nair2010rectified} for activation, which returns max($x$, 0) when passed $x$. We also apply dropout \citep{JMLR:v15:srivastava14a} after each activation, to help reduce over-fitting, with a dropout rate of 0.2 to randomly set the output of neurons to zero 20\% of the time during training. The loss of this network is determined using softmax cross entropy and optimised using the Adam algorithm \citep{2014arXiv1412.6980K} with a learning rate of $5\times10^{-5}$. The architecture can be seen schematically in Fig. \ref{fig:cnn}.

The inputs to the network are 64$\times$64 pixel images with either 3 (CANDELS) or 1 (KiDS) colour channels, that are globally scaled between 0 and 1, preserving the relative flux densities for multi colour images. The output layer has two neurons, one for each of the merging and non-merging classes, and uses a softmax output, providing the probability for each class in the range [0, 1] that sum to unity, i.e. softmax maps the un-normalised input into it to a probability distribution over the output classes if the training, test and validation data sets have an equal number of each class. In the following, we will use the output for the merger class (\texttt{frac\_merger}) although this can be considered to be equivalent to using the output for the non-merger class as it is 1-(\texttt{frac\_merger}) in our binary classification.

Due to the limitations of this architecture, specifically the use of fully connected layers, it is not possible to have input images of different pixel sizes. As a result, it is not possible to use cutouts of different sizes that maintain the relative size of the galaxy within the image. It is possible to resize the images but this risks losing small scale structure when downscaling galaxies or creating artefacts when upscaling. This may cause issues for large galaxies at low redshifts and low resolution as the primary galaxy may fill the image. However, it is possible to correctly identify the galaxies that fill, or are larger than, the image if the training set contains these types of galaxies, as has been shown in \citet{2019A&A...626A..49P}.

\subsubsection{Training, validation and testing}
If there are an unequal number of images in the two classes, the larger class size is reduced by randomly removing images until the classes are the same size. The images were then subdivided into three groups: 80\% were used for training, 10\% for validation and 10\% for testing. The training set was the set used to train the network while the validation was used to see how well the network was performing as training progressed. The testing set was used once, and once only, to test the performance of the network deemed to be the best from the validation. Testing images are not used for validation to prevent accidental training on the test data set.

During training, the input images were augmented. To increase the number of galaxies to train with for CANDELS and KiDS, the images were randomly augmented in one of the following ways: the image was translated by up to 4 pixels in each direction, zoomed by up to 2$\times$, rotated by a random angle about the galaxy, skewed by a random angle between 10$^{\circ}$ and 30$^{\circ}$ or passed without change. For the first four of these options, random Gaussian noise was added so the galaxy appeared ``new'' to the network. These augmented images were then randomly rotated by 0$^{\circ}$, 90$^{\circ}$, 180$^{\circ}$ or 270$^{\circ}$ to reduce sensitivity to galaxy orientation. Once complete, we have one network for the CANDELS data, one for the KiDS-z00 data, one for KiDS-z15, one for KiDS-z30 and one for KiDS-z45 data.

\subsection{Forward modelling of the galaxy main sequence}\label{subsec:fm}
In determining if galaxy mergers have an effect on the star-formation rates (SFR) of galaxies, it is necessary to remove the mass dependence from the star-formation rate. To do this we correct for the slope of the main sequence of star forming galaxies \citep[MS, e.g.][]{2004MNRAS.351.1151B, 2007ApJ...660L..43N, 2007A&A...468...33E, 2014ApJS..214...15S, 2018A&A...615A.146P}. The MS is an observed tight correlation between the M$_{\star}$ and SFR of star forming galaxies that exists out to at least $z = 6$ with a scatter of $\sim$0.3~dex that is mass and redshift independent. The slope of the MS is found to be less than one and to depend on the redshift \citep{2018A&A...615A.146P}. To correct for this slope, it is necessary to first determine the MS of the galaxies.

To determine the MS, we follow \citet{2018A&A...615A.146P} and use forward modelling. Assuming a linear MS, we use the Markov Chain Monte-Carlo (MCMC) sampler \texttt{emcee} \citep{2013PASP..125..306F} to simultaneously fit the slope, normalisation and scatter of all the galaxies that are star forming. At each sampled point in the parameter space, model SFRs are created using the observed stellar masses, observed redshifts and the corresponding positions along the sampled MS. These SFRs are then perturbed by selecting a random number from a Gaussian distribution, with the standard deviation equal to the sampled scatter, truncated to match the observed upper and lower SFR limits. To include observational uncertainties on SFR and M$_{\star}$, both the simulated SFR and the observed M$_{\star}$ of each galaxy are perturbed again by a random number sampled from a Gaussian distribution with the standard deviation equal to the error in the observed SFR or M$_{\star}$.

At each step, the mock SFR-M$_{\star}$ plane is compared to the observed SFR-M$_{\star}$ plane by binning the data into identical bins in M$_{\star}$. The means and standard deviations of the SFRs inside each bin are calculated and the results from the mock data are compared to the results from the observed data. The closer the means and standard deviations in the mock bins are to the observed bins, the more likely the model is a correct representation of the observed data. Due to the large number of objects, for the SDSS data we randomly select 50\,000 star forming objects within each redshift bin to determine the MS.

\section{Results}\label{sec:results}
Here we present the results of our analysis of the three data sets.

\subsection{Performance of the CNN}
The network we use for the SDSS objects is that of \citet{2019A&A...626A..49P}. For completeness, we repeat these results here. The SDSS network achieves an accuracy of 0.932 at validation with a cut threshold of 0.5, that is any object with \texttt{frac\_merger} > 0.5 is classified as a merging galaxy. We can alter the threshold value to find the threshold value that simultaneously minimises the fall out and maximises the recall. By doing this to set the threshold to 0.57, the accuracy increases to 0.935. Using the same threshold of 0.57 at testing, the final accuracy of the SDSS network is 0.915, with recall, precision, specificity and negative predictive value (NPV) of 0.920, 0.911, 0.910 and 0.919 respectively. We apply the network to 256\,497 SDSS galaxies with spectroscopic redshifts between 0.005 and 0.1, to match the training set, resulting in 28\,971$\pm$2578 (14.1$\pm$1.3\%) galaxies being identified as mergers. The errors in these merger counts are derived from the precision of the network, see Appendix \ref{app:definitions} for the definition. The number of merging galaxies is multiplied by the precision and the difference between this value and the original count is taken as the error for the number of mergers and number of non-mergers. This is likely an underestimate as the precision assumes equal population sizes of mergers and non-mergers, which is evidently not the case.

For the KiDS network, we use the CNN to identify galaxies that fall within the GAMA09 field. This network achieves an accuracy of 0.942 at validation with a cut threshold of 0.5. If we alter the threshold to 0.52, to simultaneously minimise the fall out and maximise the recall, the accuracy increases to 0.948. Using the same threshold of 0.52 at testing, the final accuracy of the KiDS network is 0.903, with recall, precision, specificity and NPV of 0.942, 0.874, 0.864 and 0.938 respectively. Despite the resolution of the KiDS images being higher than that of the SDSS images, the same image size resulted in the best performance for the KiDS network: 64$\times$64 pixels. Larger images were tried but these networks did not perform as well. Applying the KiDS network to all galaxies in the GAMA09 field with photometric redshifts below 0.15, a total of 1270 galaxies, we identify 436$\pm$55 (30.0$\pm$4.3\%) merging galaxies.

The CANDELS network achieves an accuracy of 0.826 at validation with a cut threshold of 0.5. If we decrease the threshold to 0.47, the accuracy increases to 0.840. Using the same threshold of 0.47 at testing, the final accuracy of the CANDELS network is 0.818, with recall, precision, specificity and NPV of 0.870, 0.789, 0.768 and 0.855 respectively. The poorer results for the CANDELS network is likely due to fewer pre-classified objects to train the network with, 694 per class for CANDELS compared to 3003 for SDSS, as well as the higher redshifts of the training objects. The CANDELS images also cover a much larger redshift range, resulting in a greater distribution of sizes in the image for galaxies at the same mass than the SDSS images. Ideally, it would be preferable to split the galaxies into redshift bins and train a network per redshift to minimise this effect, however with so few objects it is not feasible. We apply the CANDELS network to the objects with H-magnitude < 24.5 and $0.0 < z < 4.0$ in the CANDELS COSMOS, EGS and UDS fields and identify 3535$\pm$746 merger candidates out of the 10\,027 galaxies in these three fields. This is a merger fraction of 35.3$\pm$7.4\%, which is high. The statistics for all the networks are presented in Table \ref{table:stats:cnn} and examples of non-merger and mergers selected by the CNNs can be found in Appendix \ref{app:ex}.

\begin{table*}
	\caption{Statistics for the trained CNNs. Definitions of terms can be found in Appendix \ref{app:definitions}}
	\begin{center}
		\begin{tabular}{ccccccc}
		\hline
		 & SDSS\tablefootmark{a} & KiDS & CANDELS \\
		 & $0.005 < z < 0.1$ & $0.00 < z \leq 0.15$ & $0.00 < z \leq 4.0$ \\
		\hline
		Cut threshold &   0.57 &   0.52 & 0.47\\
		ROC area      & 0.966 & 0.957 & 0.861\\
		Recall            & 0.920 & 0.942 & 0.870\\
		Precision       & 0.911 &  0.874 & 0.789\\
		Specificity      & 0.910 & 0.864 & 0.768\\
		NPV               & 0.919 & 0.938 & 0.855\\
		Accuracy        & 0.915 & 0.903 & 0.818\\
		\hline
		\end{tabular}
	\label{table:stats:cnn}
	\tablefoot{
		\tablefoottext{a}{The SDSS network is that of \citet{2019A&A...626A..49P}.}
	}
	\end{center}
\end{table*}

\subsection{SDSS}\label{sec:sdss}
\begin{figure}
	\centering
	\includegraphics[width=0.5\textwidth]{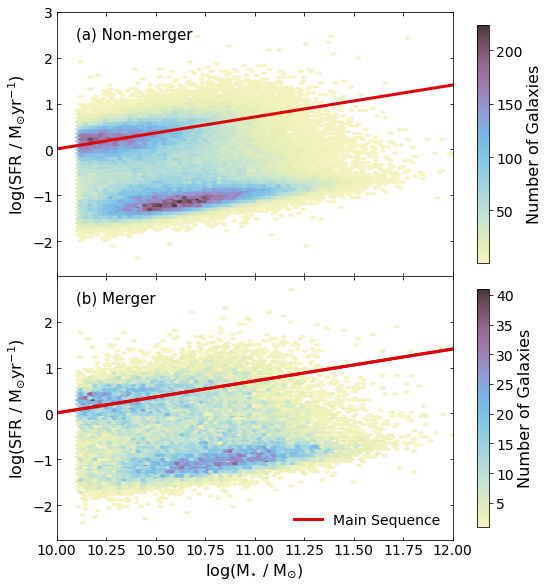}
	\caption{The SFR-M$_{\star}$ plane populated with (a) the non-merging galaxies and (b) the merging SDSS galaxies. The colour indicates the number density from low (light yellow) to high (dark purple). Overlaid in red is the MS that has been fitted to all star forming galaxies. As can be seen, the distributions of the merging and non-merging galaxies are similar with respect to the plotted MS.}
	\label{fig:sdss-SFR-Mstar}
\end{figure}

To determine the effect of galaxy mergers on SFR, we determine the effect of mergers on the MS subtracted SFR. We fit the MS to all the star forming galaxies, both mergers and non-mergers together, and the MS we have fitted to the SDSS data is shown overlaid onto all the non-merging and merging galaxies in Fig. \ref{fig:sdss-SFR-Mstar}. The MS subtracted SFR of the merging and non-merging galaxies are then compared by fitting a skewed Gaussian distribution, of the form
\begin{equation}
	y = \frac{A}{\sigma} exp\bigg(\frac{(x-\mu)}{2\sigma^{2}}\bigg) \bigg(1 + erf\bigg[\frac{\alpha(x-\mu)}{\sqrt{2}\sigma}\bigg]\bigg),
\end{equation}
to the distributions of the merging and non-merging galaxies, where $A$ is the amplitude, $\mu$ and $\sigma$ are the mean and standard deviation of the Gaussian, $\alpha$ is the description of skewness and $erf$ is the error function.

\begin{figure}
	\centering
	\includegraphics[width=0.5\textwidth]{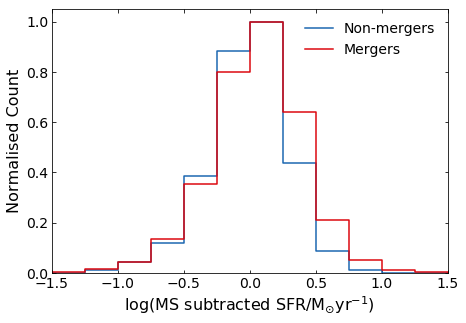}
	\caption{The distribution of MS subtracted SFR for the star forming SDSS non-merging galaxies (blue) and merging galaxies (red). As can be seen, the merging star forming population has a slightly higher mean MS subtracted SFR.}
	\label{fig:sdss-MSsub}
\end{figure}

To fit the skewed Gaussian we bin the MS subtracted SFR with bin sizes of 0.25~dex, between -3~log(M$_{\odot}$yr$^{-1}$) and 2~log(M$_{\odot}$yr$^{-1}$) and fit the skewed Gaussian to the number of galaxies in each bin. The errors on these counts were determined by generating 100 realisations of the MS subtracted SFR by perturbing the SFR and M$_{\star}$ of each galaxy by a random number drawn from a Gaussian distribution centred on the observed SFR or M$_{\star}$ and with the error on the value as the standard deviation. Each realisation was then binned in the same way as the observations and the standard deviation of the counts in the bins of the 100 realisations were taken as the errors on the counts of the observations. The \texttt{scipy.optimize} package \texttt{curve\_fit} was then used to fit the skewed Gaussian distribution to the counts in the bins and account for their errors. The distributions are presented in Fig. \ref{fig:sdss-MSsub} with the parameters for the skewed Gaussian fits in Table \ref{table:sdss-best}.

\begin{table}
	\caption{Best fit parameters for the skewed Gaussian distribution fitted to the star forming SDSS data. $\mu$ and $\sigma$ are in units of log(M$_{\odot}$yr$^{-1}$)}
	\begin{center}
		\begin{tabular}{ccc}
		\hline
		Parameter & Merger & Non-merger \\
		\hline
		$\mu$ & 0.33 $\pm$ 0.02 & 0.25 $\pm$ 0.01 \\
		$\sigma$ & 0.43 $\pm$ 0.02 & 0.39 $\pm$ 0.01 \\
		$\alpha$ & -1.29 $\pm$ 0.18 & -1.52 $\pm$ 0.16 \\
		\hline
		\end{tabular}
	\label{table:sdss-best}
	\end{center}
\end{table}

Comparing the skewed Gaussian fits to the distributions, we find that the mean for the star forming mergers and non-mergers are consistent within 3 times the error of the mean ($\sigma_{\mu}$) and the merging galaxies have higher mean MS subtracted SFR. This suggests that the star forming population has a slightly, but not significantly, increased SFR when undergoing a merger.

\subsection{KiDS}
\begin{figure}
	\centering
	\includegraphics[width=0.5\textwidth]{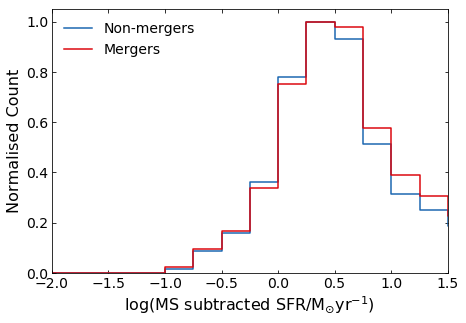}
	\caption{The distribution of MS subtracted SFR for the star forming KiDS non-merging galaxies (blue) and merging galaxies (red). As can be seen, the merging star forming galaxies have a similar mean MS subtracted SFR to the non-merging galaxies.}
	\label{fig:kids-MSsub}
\end{figure}

As with the SDSS data, we fit a skewed Gaussian distribution to the MS subtracted SFR of the star forming galaxies. An example of the resulting MS subtracted SFR distributions for the KiDS is shown in Fig. \ref{fig:kids-MSsub}. Table \ref{table:kids-best} shows that the merging star forming galaxies have higher average SFRs. The differences not large, with the mean MS subtracted SFR being within 3$\sigma_{\mu}$ of each other.

\begin{table}
	\caption{Best fit parameters for the skewed Gaussian distribution fitted to the star forming KiDS data. $\mu$ and $\sigma$ are in units of log(M$_{\odot}$yr$^{-1}$)}
	\begin{center}
		\begin{tabular}{cccc}
		\hline
		Parameter & Merger & Non-merger \\
		\hline
		
		$\mu$ & 0.44 $\pm$ 0.1 & 0.13 $\pm$ 0.18 \\
		$\sigma$ & 0.47 $\pm$ 0.09 & 0.41 $\pm$ 0.08 \\
		$\alpha$ & -1.53 $\pm$ 0.91 & -0.63 $\pm$ 0.78 \\
		\hline
								
		\end{tabular}
	\label{table:kids-best}
	\end{center}
\end{table}

\subsection{CANDELS}
Due to the larger redshift coverage of the CANDELS data, we can examine if the impact of galaxy mergers on SFR changes as a function of redshift. To do this, we divided the data into redshift bins with edges at $z =$ 0.0, 0.6, 0.85, 1.21, 1.66 and 4.0, each with its own mass completeness limit and containing approximately 2000 galaxies after mass completeness cuts have been applied. Each redshift bin also had its own main sequence fitted as outlined in Sect. \ref{subsec:fm}. For ease of reference, these redshift bins shall be referred to as CANDELS-z000, CANDELS-z060, CANDELS-z085, CANDELS-z121 and CANDELS-z166.

\begin{figure}
	\centering
	\includegraphics[width=0.5\textwidth]{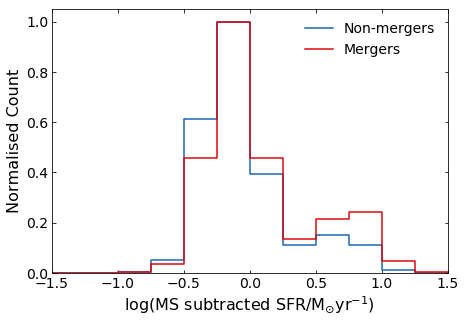}
	\caption{The distribution of MS subtracted SFR for the $0.85 < z \leq 1.21$ redshift bin for the CANDELS non-merging galaxies (blue) and merging galaxies (red). This is the only data that is fitted with a double Gaussian distribution due to the clear multi-modal population. As can be seen, the main and secondary populations have a slightly higher mean MS subtracted SFR than the non-merging galaxies.}
	\label{fig:candels-MSsub}
\end{figure}

As before, we fit the distributions of the star forming CANDELS galaxies with a skewed Gaussian function. However, there is an indication of a second, high SFR population in CANDELS-z085 ($0.85 < z \leq 1.21$), identifiable when the error on the skew of the both the merging and non-merging distributions are greater than $10^{4}$ and so in that bin only, we fit a double Gaussian distribution and consider the lower mean to be the mean of the star forming population. The distributions for the MS subtracted SFR for this redshift bin is shown in Fig. \ref{fig:candels-MSsub}.

Using the best fitting values for the skewed and double Gaussian functions presented in Table \ref{table:candels-best}, in the lowest redshift bin the merging galaxies act to suppress the SFR of galaxies, with the mean MS subtracted SFR for merging galaxies lower than non-merging galaxies by more than 3$\sigma_{\mu}$. At redshifts above $z = 0.60$, the star forming mergers have a higher mean than the star forming non-mergers or are consistent within 3$\sigma_{\mu}$.

\begin{table*}
	\caption{Best fit parameters for the skewed or double Gaussian distributions fitted to the star forming CANDELS data. For the $0.85 < z \leq 1.21$ bin, where a double Gaussian is used, the star forming component is the component with the lowest $\mu$. $\mu$ and $\sigma$ are in units of log(M$_{\odot}$yr$^{-1}$)}
	\begin{center}
		\begin{tabular}{cccc}
		\hline
		Redshift & Parameter & Merger & Non-merger \\
		\hline
		
		\multirow{3}{*}{$ 0.0 < z \leq 0.6 $} & $\mu$ & -0.37 $\pm$ 0.17 & -0.27 $\pm$ 0.02 \\
		 & $\sigma$ & 0.47 $\pm$ 0.08 & 0.29 $\pm$ 0.02 \\
		 & $\alpha$ & 9999.489 $\pm$ 87523220.85 & 1.21 $\pm$ 0.26 \\
		\hline
		
		\multirow{3}{*}{$ 0.6 < z \leq 0.85 $} & $\mu$ & 0.10 $\pm$ 0.12 & 0.02 $\pm$ 0.11 \\
		 & $\sigma$ & 0.24 $\pm$ 0.07 & 0.24 $\pm$ 0.06 \\
		 & $\alpha$ & -0.878 $\pm$ 1.17 & -0.79 $\pm$ 0.99 \\
		\hline
		
		\multirow{2}{*}{$ 0.85 < z \leq 1.21 $} & $\mu$ & -0.12 $\pm$ 0.01 & -0.16 $\pm$ 0.01 \\
		 & $\sigma$ & 0.20 $\pm$ 0.01 & 0.20 $\pm$ 0.01 \\
		\hline
		
		\multirow{3}{*}{$ 1.21 < z \leq 1.66 $} & $\mu$ & -0.43 $\pm$ 0.02 & -0.42 $\pm$ 0.02 \\
		 & $\sigma$ & 0.62 $\pm$ 0.03 & 0.52 $\pm$ 0.03 \\
		 & $\alpha$ & 5.058 $\pm$ 1.24 & 3.13 $\pm$ 0.69 \\
		\hline
		
		\multirow{3}{*}{$ 1.66 < z \leq 4.0 $} & $\mu$ & -0.47 $\pm$ 0.02 & -0.5 $\pm$ 0.02 \\
		 & $\sigma$ & 0.65 $\pm$ 0.03 & 0.61 $\pm$ 0.03 \\
		 & $\alpha$ & 3.241 $\pm$ 0.51 & 3.7 $\pm$ 0.61 \\
		\hline
				
		\end{tabular}
	\label{table:candels-best}
	\end{center}
\end{table*}

\section{Discussion}\label{sec:discussion}
Here we present discussions of our results. We note that direct comparisons between the results of the three data sets is difficult due to the different definitions of mergers employed for the training data sets as well as difference in data quality, such as depth and resolution, which can also influence merger identification. While the merger definitions are similar, as they are all based on visual classification, the specifics of the definitions differ. The classifications also cover both major and minor mergers, with approximately half of each training set comprising of major mergers. This likely results in a similar split for the mergers classified by our networks.

\subsection{Merger influence on SFR}
Across the SDSS, CANDELS and KiDS data sets there is a difference between the SFRs of the merging and non-merging galaxies. However, the difference between the two is small and varies between the data sets as well as within the data sets. What is evident is that the merging systems are not only found as starburst galaxies but also as star forming and quiescent systems. 

\begin{figure}
	\centering
	\includegraphics[width=0.5\textwidth]{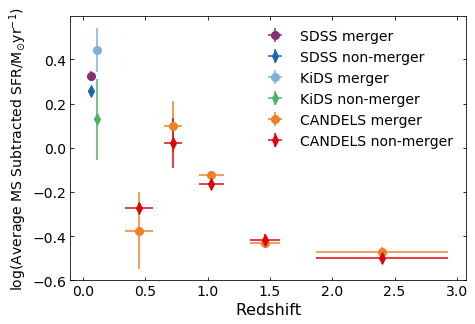}
	\caption{The average MS subtracted SFR  of the star forming galaxies for the SDSS merging (purple circle) and non-merging (dark blue diamond); KiDS merging (light blue circle) and non-merging (green diamond); and CANDELS merging (orange circles) and non-merging (red diamonds) galaxies. As can be seen, the change in SFR between the merging and non-merging galaxies is typically small.}
	\label{fig:summary}
\end{figure}

Comparing the SDSS data with the KiDS data, we find little difference in how mergers are affecting the SFR. Both data sets show that star forming merging galaxies have a slight increase in SFR. Within the CANDELS-z000 data the opposite is found: we find that there is a decrease in the MS subtracted SFR, suggesting that galaxy mergers are acting to reduce the SFR of the star forming galaxies. The full comparison between the average SFRs for all data sets at all redshifts studied can be seen in Fig. \ref{fig:summary}.

The slight difference between the merging and non-merging SFRs is also not a result of the observation bands or methods used to derive the SFR. The mergers in all three surveys are detected using different bands: SDSS uses three optical bands (gri), CANDELS uses observed frame near infrared (Hubble 1.6~$\mu$m, 1.25~$\mu$m and 814~nm bands) and KiDS uses a single optical band (r band). SFRs in the three surveys are also derived differently: SDSS uses H$\alpha$ based SFR while CANDELS and KiDS use SED derived SFR. The models used to derive the CANDELS and KiDS SFRs and M$_{\star}$ are also different. Thus, the small effect of merging galaxies on the SFRs seen is this study is robust.

Our results are qualitatively in line with previous work in that we only find small (less than a factor of two) changes in SFR. \citet{2014AJ....148..137L} and \citet{2015MNRAS.454.1742K} find that mergers change the SFR by up to a factor of two. While we do not find that mergers always result in an increase in SFR, we do find that the change in SFR caused by a galaxy merger is typically small over the timescale of the entire merger. If an increase in SFR due to a galaxy merger is large but shorter lived, the effect will be hidden by the larger number of galaxies not undergoing such a burst of star formation. The changes in average MS subtracted SFR are small and typically found to be a factor of $\sim$1.2. Similarly, \citet{2018ApJ...868...46S} find that mergers produce no significant change to the SFR of galaxies, which is consistent with the results of our study. However, caution must be taken with this comparison as the work of \citet{2018ApJ...868...46S} uses mergers where the two merging galaxies are within 3-15~kpc of each other, something that this work does not take into account. 

This study has its limitations. It is likely that we are observing different stages of galaxy mergers but our method is currently unable to determine at what stage the mergers are. As a result, it is not possible to say, from this study, if mergers cause a migrating of the merging galaxies across the SFR-M$_{\star}$ plane or if the merger only slightly affects the SFR resulting in the small changes we observe.

\subsection{Merger fractions}
\begin{figure}
	\centering
	\includegraphics[width=0.5\textwidth]{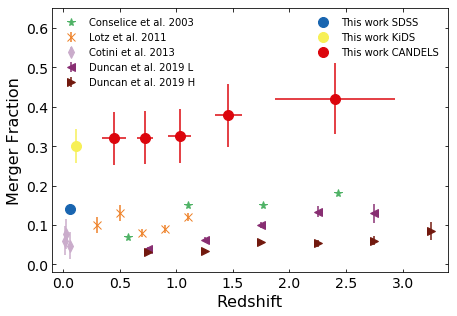}
	\caption{The total merger fraction as a function of redshift for the SDSS (dark blue circle), KiDS (light blue circle) and CANDELS (red circles) by redshift bin. Also plotted are the mass limited merger fractions with log(M$_{\star}$/M$_{\odot}$) > 10.0 from \citet[][green stars]{2003AJ....126.1183C}, \citet[][lilac diamonds]{2013MNRAS.431.2661C}, \citet{2011ApJ...742..103L} magnitude limited merger fractions with M$_{B} >$ -19.2 (orange crosses) and the \citet{2019ApJ...876..110D} lower mass (9.7 < log(M$_{\star}$/M$_{\odot}$ < 10.3, L, purple left triangles) and higher mass (log(M$_{\star}$/M$_{\odot}$ > 10.3, H, brown right triangles) merger fractions. The SDSS data are slightly higher than would be expected and the KiDS and CANDELS merger fractions are approximately a factor of two higher than the other results.}
	\label{fig:mergerfrac-comp}
\end{figure}

The merger fractions for CANDELS, 35.3$\pm$7.4\%, and KiDS, 36.9$\pm$5.3\%, are notably higher than the merger fractions for the SDSS data at 14.1$\pm$1.3\%, see Table \ref{table:mergerfrac} and Figs. \ref{fig:mergerfrac-comp} and \ref{fig:mergerfrac-tw}. The errors in these merger fractions are derived from the precision of the network, see Appendix \ref{app:definitions} for the definition. The number of merging galaxies is multiplied by the precision and the difference between this value and the original count is taken as the error for the number of mergers and number of non-mergers. The merging fraction for the precision corrected counts is then calculated and the difference between the original fraction and this precision corrected fraction is taken as the error. This is likely an underestimate as the precision assumes equal population sizes of mergers and non-mergers, which is evidently not the case.

\begin{table}
	\caption{The merger fraction by redshift and data set for the quiescent, star forming and total galaxy populations. Errors are derived from correcting for the precision of the network.}
	\begin{center}
		\begin{tabular}{cccc}
		\hline
		Data Set & Total & Quiescent & Star forming\\
		\hline
		SDSS & 14.1$\pm$1.3\% & 14.3$\pm$1.3\% & 13.4$\pm$1.2\% \\
		\hline
		KiDS& 30.0$\pm$4.3\% & 19.4$\pm$2.8\% & 36.9$\pm$5.3\% \\
		\hline
		CANDELS-z000 & 32.0$\pm$6.8\% & 30.0$\pm$6.2\% & 32.4$\pm$6.8\% \\
		CANDELS-z060 & 32.2$\pm$6.8\% & 20.2$\pm$4.3\% & 33.6$\pm$7.1\% \\
		CANDELS-z085 & 32.6$\pm$6.9\% & 24.4$\pm$5.1\% & 33.3$\pm$7.0\% \\
		CANDELS-z121 & 37.8$\pm$8.0\% & 23.9$\pm$5.3\% & 39.4$\pm$8.3\% \\
		CANDELS-z166 & 42.1$\pm$8.9\% & 28.5$\pm$5.9\% & 44.3$\pm$9.3\% \\
		\hline
		\end{tabular}
	\label{table:mergerfrac}
	\end{center}
\end{table}

Even only considering the lowest redshift bin for CANDELS, $0.00 < z \leq 0.60$, the merger fraction is much higher than the SDSS and higher redshift KiDS at 32.0$\pm$6.8\%. It is unsurprising that this becomes a larger issue as the redshift increases because the pixel size of the galaxy within the image becomes smaller and the galaxies themselves become fainter, suppressing the features that the CNN will look for to identify a merging galaxy.

Comparing our merger fractions to other works shows that the CANDELS results are indeed much higher than would be expected. Figure \ref{fig:mergerfrac-comp} shows the comparison of this work with \citet{2003AJ....126.1183C}, who use CAS to identify mergers, \citet{2011ApJ...742..103L}, who use Gini and $M_{20}$, \citep{2013MNRAS.431.2661C}, who use CAS, Gini and $M_{20}$, and \citet{2019ApJ...876..110D}, who use the close pair method. The results of \citet{2019ApJ...876..110D} are the merger pair fraction (the number of pairs of merging galaxies divided by the total number of galaxies) and so we multiply their values by 0.6 to compare to our results \citep{2011ApJ...742..103L, 2017MNRAS.470.3507M}.

The SDSS merger fraction is higher than the other works in the same redshift range but is consistent with the merger fractions of \citet{2003AJ....126.1183C} and \citet{2011ApJ...742..103L} at higher redshifts. The KiDS data has a merger fraction that is higher compared to the other works, both at similar and higher redshifts, similar to the merger fractions from CANDELS as discussed above.

We can compare the merger fractions of the quiescent and star forming galaxies as shown in Fig. \ref{fig:mergerfrac-tw}. The SDSS data has a slightly lower merger fraction for the quiescent galaxies than the star forming galaxies, although the difference is 0.2 percentage points, much less than the error on the merger fractions. KiDS data has a higher merger fraction for the quiescent galaxies than the star forming galaxies. As these two data sets cover similar redshift ranges one would expect to see a similar trend in the merger fractions of these two populations. The difference in overall merger fractions may be a result of the SDSS and KiDS networks not being identical and the different selection criteria for the training sets.

\begin{figure}
	\centering
	\includegraphics[width=0.5\textwidth]{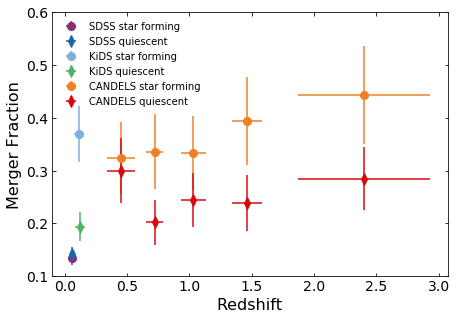}
	\caption{The merger fraction of quiescent (diamonds) and star forming (circles) as a function of redshift for SDSS (purple and blue), KiDS (yellow and green) and CANDELS (orange and red). There is no overall trend with redshift with SDSS having a lower merger fraction for the star forming galaxies, KiDS having a higher merger fraction for star forming galaxies and CANDELS star forming galaxies having a higher merger fraction at all redshifts.}
	\label{fig:mergerfrac-tw}
\end{figure}

This is qualitatively different to the CANDELS-z000 data that has a slightly lower quiescent merger fraction than star forming merger fraction. This difference is more pronounced at higher redshifts resulting in a different conclusion from the KiDS data. The CANDELS data suggests that there is a higher fraction of star forming galaxy mergers than quiescent galaxy mergers at all redshifts, implying that galaxy mergers do not often act to suppress star formation rates.

The CNNs used in this work are not perfect as they misclassify mergers as non-mergers and non-mergers as mergers. The latter of these misclassifications may present issues with our analysis. As non-mergers are more prevalent than mergers, relatively high specificity of a network can still result in a large population of non-merging galaxies being added to the merging classification. If galaxy mergers do significantly change the SFR of the galaxies, the non-merging interlopers may act to suppress this effect in the statistical analysis used in this paper. However, as this work is primarily comparing the relative SFRs of merging and non-merging galaxies, we do not believe that this overly impacts our results as the differences we see in star formation rates between the mergers and non-mergers is small.

\subsection{Starburst merger fraction}
\begin{figure}
	\centering
	\includegraphics[width=0.5\textwidth]{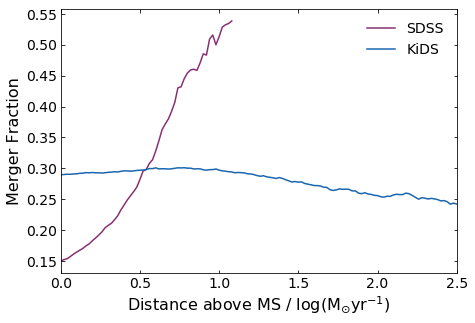}
	\includegraphics[width=0.5\textwidth]{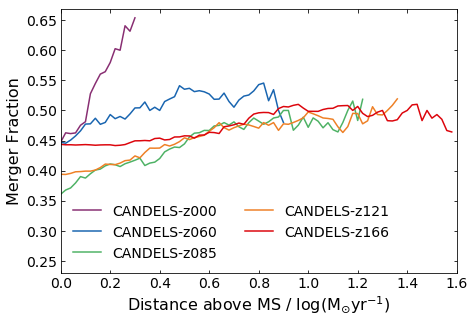}
	\caption{Merger fraction for star forming galaxies with SFRs above the indicated distance above the MS for the SDSS and KiDS data (top panel) and the CANDELS data (bottom panel). To avoid low number statistics, only thresholds above which there are 50, or more, galaxies are shown. The SDSS (top panel, purple), KiDS (top panel, blue) and all CANDELS data show a trend of increasing merger fraction as the distance to the MS increases, although the CANDELS-z000 drops again above 0.62~log(M$_{\odot}$yr$^{-1}$).}
	\label{fig:SB-merger}
\end{figure}

We avoid using a specific definition of a starburst galaxy and instead opt to study the merger fraction as a function of distance above the MS. For ease of reference, we will refer to the galaxies above a given SFR threshold as starbursting in this subsection, even if the threshold is the MS. To this end, we study the fraction of star forming galaxies above a certain distance above the MS that are merging for all three data sets (number of merging galaxies above a certain threshold / total number of galaxies above the same threshold).
 These trends are presented in Fig. \ref{fig:SB-merger}.

The SDSS and KiDS data show an increase in the merger fraction as the distance from the MS increases, with SDSS rising to $\sim$1.1~log(M$_{\odot}$yr$^{-1}$) and KiDS slowly declining above $\sim$0.8~log(M$_{\odot}$yr$^{-1}$). A similar trend is seen in the CANDELS data, with an increase in merger fraction as the distance from the MS increases. CANDELS-z000 rises to approximately 0.3~log(M$_{\odot}$yr$^{-1}$) while the other three CANDELS redshift bins rise to approximately 0.8, 1.2, 1.4 and 1.6 for CANDELS-z060, CANDELS-z085, CANDELS-z121 and CANDELS-z166 respectively. Thus, the merger fraction increases as the star formation rate increases showing that mergers can act to trigger high star formation rates and starbursts. We note, however, that the number of galaxies in Fig. \ref{fig:SB-merger} decreases as the distance above the MS increases meaning that the lower merger fraction at lower distance can contain more mergers than the higher merger fraction at larger distances. This allows for the small changes in SFR seen in the star forming population despite the merger fraction increasing as the distance above the MS increases. This is qualitatively consistent with \citep{2014ApJ...789L..16L}, who find approximately half of starburst systems (defined as an increase in SFR by a factor of 5 or more) are undergoing a merger while the fraction of mergers in non-starburst systems is lower.

\begin{figure}
	\centering
	\includegraphics[width=0.5\textwidth]{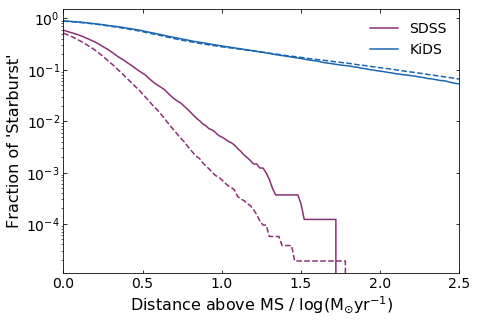}
	\includegraphics[width=0.5\textwidth]{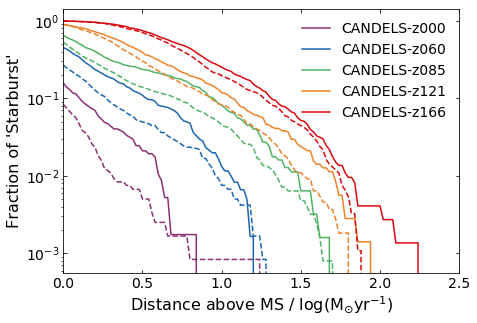}
	\caption{Fraction of star forming, merging galaxies with SFRs above a given distance above the MS (solid lines) and the fraction of star forming, non-merging galaxies with SFRs above a given distance above the MS (dashed lines). The top panel contains the SDSS and KiDS data sets while the bottom panel contains the CANDELS data. To avoid low number statistics, only thresholds above which there are 50, or more, galaxies are shown. The SDSS (top panel, purple), KiDS (top panel, blue) and all CANDELS data show that there is a higher fraction of the total number of merging galaxies above nearly all distance above the MS.}
	\label{fig:SB-merger-inv}
\end{figure}

We can also compare the fraction of star forming, merging galaxies that have SFRs above a certain distance above the MS (number of merging galaxies above a certain threshold / total number of merging galaxies) with the fraction of star forming, non-merging galaxies that have SFRs above a certain distance above the MS as shown in Fig. \ref{fig:SB-merger-inv}. The SDSS, KiDS and CANDELS data all show that a higher fraction of starburst mergers are found than the fraction of starburst non-mergers, although this switches at larger distances for the KiDS data. This is clear evidence that the merging galaxies are causing an increase in SFR.

\section{Conclusions}\label{sec:conclusion}
Galaxy mergers are an important part of how galaxies grow and evolve over the history of the universe. However, identifying galaxy mergers is a difficult and time-consuming task. Here we have employed deep learning techniques to identify galaxy mergers in SDSS, KiDS and CANDELS imaging data. We have then used these classifications to explore how galaxy mergers affect SFRs.

We find that mergers do indeed influence the SFR in the merging galaxies. However, the resulting change in SFR is small, typically a factor of $\sim$1.2. Within the SDSS data, the star forming objects have a slight increase, on average, that is also seen in the KiDS data within a similar redshift range. Between $0.0 < z \leq 0.6$, the CANDELS data shows a slight decrease in SFR for the star forming population when examining the MS subtracted SFR. Continuing to higher redshifts with the CANDELS data, we again find slight increases SFR for the merging galaxies with.

Overall, the change seen in the SFR of the star forming population is small, with the majority of changes in the SFR in all data sets being less than 3$\sigma_{\mu}$, a factor of $\sim$1.2.

The merger fraction of quiescent and star forming galaxies also depends on the data set. The SDSS data has a slightly high merger fraction for quiescent galaxies compared to star forming galaxies while the KiDS and CANDELS data is the opposite. Again, definite conclusions are difficult with the CANDELS and KiDS data showing that galaxy mergers are more common in star forming galaxies at any redshift while the SDSS data does not.

Instead of directly examining the fraction of starburst galaxies that are mergers, we examine the merger fraction as a function of distance above the MS. For the SDSS, CANDELS and KiDS the fraction of mergers increases as the distance above the MS increases. This is evidence that mergers can cause periods of enhanced star formation.

Our current work does not determine the stage of the galaxy merger but we can see by eye that our merger samples include mergers at different stages. Thus, it is possible that the period during which SFR is boosted significantly is very short during the merging process and missed within our more time averaged analysis. It could also be that SFR is only boosted significantly for a small fraction of merger types or a combination of both scenarios. Future work will aim to overcome these shortcomings by determining the merger stage.

\begin{acknowledgements}
We would like to thank the anonymous referee for their thoughtful comments that have improved the quality of this paper.\\

We would like to thank the Center for Information Technology of the University of Groningen for their support and for providing access to the Peregrine high performance computing cluster.\\

MB is supported by the Polish Ministry of Science and Higher Education through grant DIR/WK/2018/12.\\

Funding for the SDSS and SDSS-II has been provided by the Alfred P. Sloan Foundation, the Participating Institutions, the National Science Foundation, the U.S. Department of Energy, the National Aeronautics and Space Administration, the Japanese Monbukagakusho, the Max Planck Society, and the Higher Education Funding Council for England. The SDSS Web Site is http://www.sdss.org/.\\

The SDSS is managed by the Astrophysical Research Consortium for the Participating Institutions. The Participating Institutions are the American Museum of Natural History, Astrophysical Institute Potsdam, University of Basel, University of Cambridge, Case Western Reserve University, University of Chicago, Drexel University, Fermilab, the Institute for Advanced Study, the Japan Participation Group, Johns Hopkins University, the Joint Institute for Nuclear Astrophysics, the Kavli Institute for Particle Astrophysics and Cosmology, the Korean Scientist Group, the Chinese Academy of Sciences (LAMOST), Los Alamos National Laboratory, the Max-Planck-Institute for Astronomy (MPIA), the Max-Planck-Institute for Astrophysics (MPA), New Mexico State University, Ohio State University, University of Pittsburgh, University of Portsmouth, Princeton University, the United States Naval Observatory, and the University of Washington.\\

Based on observations made with ESO Telescopes at the La Silla Paranal Observatory under programme IDs 177.A-3016, 177.A-3017, 177.A-3018 and 179.A-2004, and on data products produced by the KiDS consortium. The KiDS production team acknowledges support from: Deutsche Forschungsgemeinschaft, ERC, NOVA and NWO-M grants; Target; the University of Padova, and the University Federico II (Naples).\\

GAMA is a joint European-Australasian project based around a spectroscopic campaign using the Anglo-Australian Telescope. The GAMA input catalogue is based on data taken from the Sloan Digital Sky Survey and the UKIRT Infrared Deep Sky Survey. Complementary imaging of the GAMA regions is being obtained by a number of independent survey programmes including GALEX MIS, VST KiDS, VISTA VIKING, WISE, Herschel-ATLAS, GMRT and ASKAP providing UV to radio coverage. GAMA is funded by the STFC (UK), the ARC (Australia), the AAO, and the participating institutions. The GAMA website is http://www.gama-survey.org/ \\

This work is based on observations taken by the CANDELS Multi-Cycle Treasury Program with the NASA/ESA HST, which is operated by the Association of Universities for Research in Astronomy, Inc., under NASA contract NAS5-26555.\\

This research made use of Astropy,\footnote{\url{http://www.astropy.org}} a community-developed core Python package for Astronomy \citep{2013A&A...558A..33A, 2018AJ....156..123A}.
\end{acknowledgements}

\bibliographystyle{aa} 
\bibliography{Paper-AA-2019-36337} 

\begin{appendix}
\section{De-blending the SPIRE data}\label{app:cigale}
For de-blending the SPIRE data, we follow \citet{2017A&A...603A.102P}. CIGALE is used with the 9-band KiDS catalogue data to generate estimates of the SPIRE flux densities and we then select all objects with a predicted 250~$\mu$m flux density above 1.1~mJy. The CIGALE flux density estimates are then used as a flux density prior inside XID+ and all three SPIRE bands in the GAMA09 field are de-blended. For the CIGALE models, we follow \citet{2018A&A...615A.146P} but remove the active galactic nuclei component, due to the limited wavelength coverage available, and increase the sampling of the age of the stellar population.

For our CIGALE models, we follow \citep{2018A&A...615A.146P} but remove the active galactic nuclei component, due to the limited wavelength coverage available, and increase the sampling of the age of the stellar population. Thus, we use a double exponentially declining star formation history, \citet{2003MNRAS.344.1000B} stellar emission, \citet{2003PASP..115..763C} initial mass function (IMF), \citet{2000ApJ...539..718C} dust attenuation and the updated \citet{2014ApJ...780..172D} version of the \citet{2007ApJ...657..810D} infrared dust emission. A list of parameters, where they differ from the default values, can be found in Table \ref{table:params}.1.

\begin{table*}\label{table:params}
	\caption{Parameters used for the various properties in the CIGALE model SEDs where they differ from the default values. All ages and times are in Gyr.}
	\label{table:parameters}
	\centering
	\begin{tabular}{c c c}
		\hline
		Parameter & Value & Description\\
		\hline
		\multicolumn{3}{c}{Star Formation History} \\
		\hline
		& & \\
		$\tau_{\mathrm{main}}$ & 1.0, 1.8, 3.0, 5.0, 7.0 & e-folding time (main)\\
		$\tau_{\mathrm{burst}}$ & 9.0, 13.0 & e-folding time (burst) \\
		$f_{\mathrm{burst}}$ & 0.00, 0.10, 0.20, 0.30, 0.40 & Burst mass fraction\\
		Age & 0.50, 1.00, 1.50, 2.00, 2.50, 3.00, & Population age (main)\\
		 & 3.50, 4.00, 4.50, 5.00, 5.50, 6.00, & \\
		 & 6.50, 7.00, 7.50, 8.00, 8.50, 9.00, & \\
		 & 9.50, 10.00, 10.50, 11.00, 12.00, 13.00 & \\
		Burst Age & 0.001, 0.010, 0.030, 0.100, 0.300 & Population age (burst)\\
		& & \\
		\hline
		\multicolumn{3}{c}{Stellar Emission}\\
		\hline
		& & \\
		IMF & \citet{2003PASP..115..763C} & Initial Mass Function\\
		$Z$ & 0.02 & Metallicity (0.02 is Solar)\\
		Separation Age & 0.01 & Separation between young and old stellar populations\\
		& & \\
		\hline
		\multicolumn{3}{c}{Dust Attenuation}\\
		\hline
		& & \\
		A$_\mathrm{V}^{\mathrm{BC}}$ & 0.3, 1.2, 2.3, 3.3, 3.8 & V-band attenuation of the birth clouds\\
		Slope$_{\mathrm{BC}}$ & -0.7 & Birth cloud attenuation power law slope\\
		BC to ISM Factor & 0.3, 0.5, 0.8, 1.0 & Ratio of the birth cloud attenuation to ISM attenuation\\
		Slope$_{\mathrm{ISM}}$ & -0.7 & ISM attenuation power law slope\\
		& & \\
		\hline
		\multicolumn{3}{c}{Dust Emission}\\
		\hline
		& & \\
		$q_{\mathrm{PAH}}$ & 0.47, 1.12, 2.50, 3.9 & Mass fraction of PAH \\
		$U_{\mathrm{min}}$ & 5.0, 10.0, 25.0 & Minimum scaling factor of the radiation field intensity\\
		$\alpha$ & 2.0 & Dust power law slope\\
		$\gamma$ & 0.02 & Illuminated fraction\\
		& & \\
		\hline
	\end{tabular}
\end{table*}

\section{CNN performance definitions}\label{app:definitions}
This paper uses the definitions of \citet{2019A&A...626A..49P} for the terms to describe the properties of CNNs. These terms may be an alternate nomenclature to other works or may be unfamiliar. To avoid confusion we reproduce these definitions in Table \ref{table:definitions}.

\begin{table*}
	\caption{Terms used when describing the performance of neural networks from \citet{2019A&A...626A..49P}}
	\begin{center}
		\begin{tabular}{p{0.16\textwidth}p{0.54\textwidth}p{0.23\textwidth}}
		\hline
		Term & Definition & \\
		\hline
		Positive (P) & An object classified in the catalogues or identified by a network as a merger. & \\
		Negative (N) & An object classified in the catalogues or identified by a network as a non-merger. & \\
		True Positive (TP) & An object classified in the catalogues as a merger that is identified by a network as a merger. & \\
		False Positive (FP) & An object classified in the catalogues as a non-merger that is identified by a network as a merger. & \\
		True Negative (TN) & An object classified in the catalogues as a non-merger that is identified by a network as a non-merger. & \\
		False Negative (FN) & An object classified in the catalogues as a merger that is identified by a network as a non-merger. & \\
		Recall & Fraction of objects correctly identified by a network as a merger with respect to the total number of objects classified in the catalogues as mergers. & TP / (TP+FN) \\ 
		Fall-out & Fraction of objects incorrectly identified by a network as a merger with respect to the total number of objects classified in the catalogues as mergers. &  FP / (TP+FN) \\ 
		Specificity & Fraction of objects correctly identified by a network as a non-merger with respect to the total number of objects classified in the catalogues as non-mergers. & TN / (TN+FP) \\
		Precision & Fraction of objects correctly identified by a network as a merger with respect to the total number of objects identified by a network as a merger. & TP / (TP+FP) \\
		Negative Predictive Value (NPV) & Fraction of objects correctly identified by a network as a non-merger with respect to the total number of objects identified by a network as a non-merger. & TN / (TN+FN) \\
		Accuracy & Fraction of objects, both merger and non-merger, correctly identified by a network. & (TP+TN) / (TP+FP+TN+FN) \\
		\hline
		\end{tabular}
	\label{table:definitions}
	\end{center}
\end{table*}

\section{Example non-mergers and mergers}\label{app:ex}
\begin{figure}
	\centering
	\includegraphics[width=0.5\textwidth]{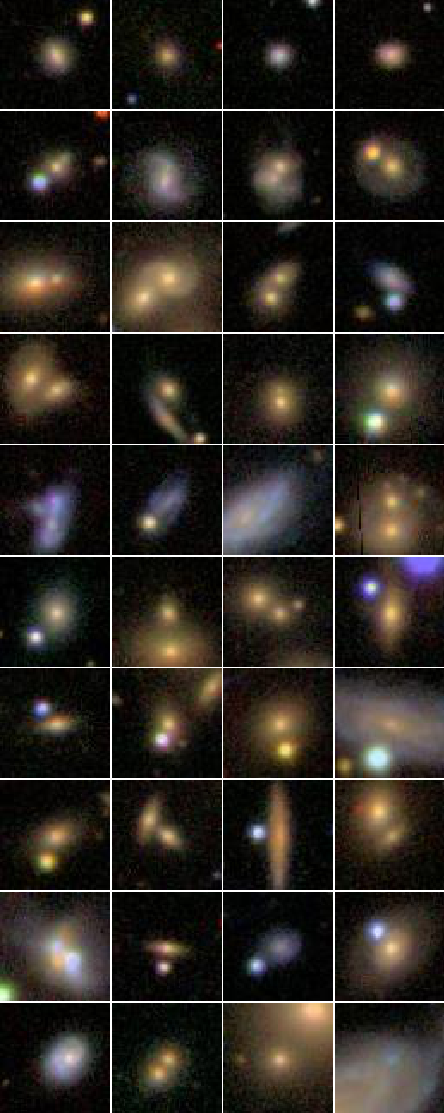}
	\caption{Examples of non-merging galaxies (top row) and merging galaxies (bottom nine rows) for the SDSS data set as defined by the CNN. Images are gri composite with a size of 64$\times$64~pixel or 13.7$\times$13.7~arcsec.}
	\label{fig:ex-sdss}
\end{figure}

\begin{figure}
	\centering
	\includegraphics[width=0.5\textwidth]{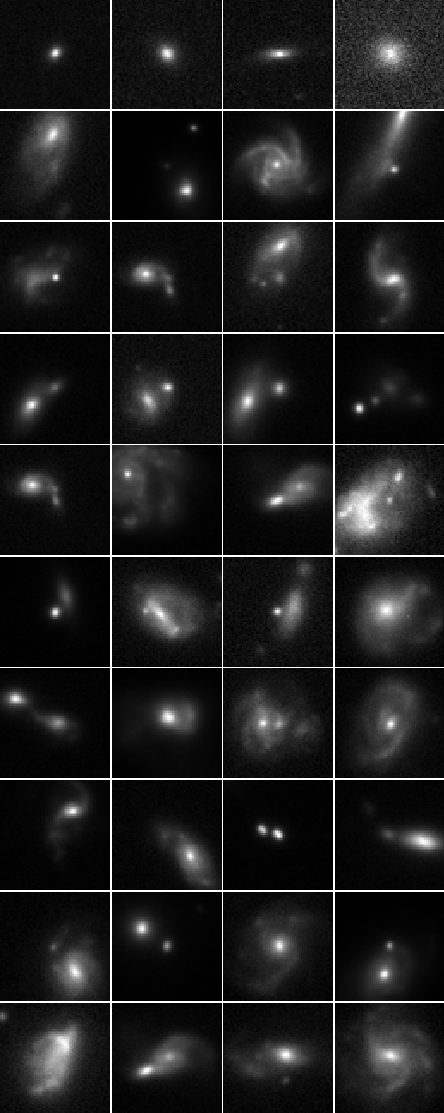}
	\caption{Examples of non-merging galaxies (top row) and merging galaxies (bottom nine rows) for the KiDS data sets as defined by the CNN. Images are grayscale r-band with a size of 64$\times$64~pixel or 25.3$\times$25.3~arcsec.}
	\label{fig:ex-kids}
\end{figure}

\begin{figure}
	\centering
	\includegraphics[width=0.5\textwidth]{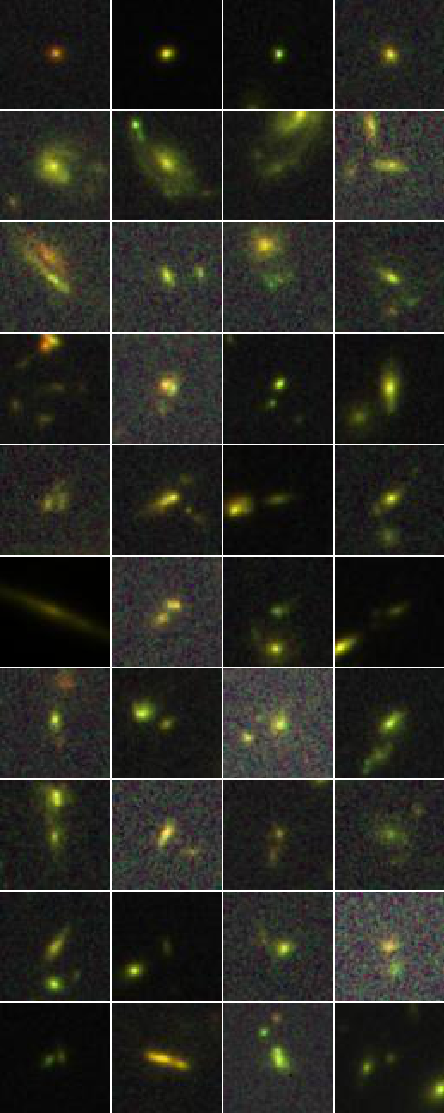}
	\caption{Examples of non-merging galaxies (top row) and merging galaxies (bottom nine rows) for the CANDELS data set as defined by the CNN. Images are 1.6~$\mu$m, 1.25~$\mu$m, 814~nm composite with a size of 64$\times$64~pixel or 3.8$\times$3.8~arcsec.}
	\label{fig:ex-candels}
\end{figure}

Here we present example non-mergers and mergers as defined by the CNN. The images shown are all 64$\times$64~pixel images with gri composite for SDSS, Fig. \ref{fig:ex-sdss}, grayscale r-band for KiDS, Fig. \ref{fig:ex-kids}, and 1.6~$\mu$m, 1.25~$\mu$m, 814~nm composite for CANDELS, Fig. \ref{fig:ex-candels}.

\end{appendix}

 \end{document}